\documentclass[twocolumn]{ceurart}

\usepackage{libertine}
\usepackage{listings}
\usepackage{xcolor}
\usepackage{multicol}
\usepackage{tikz}
\usepackage{newtxtt}
\usepackage{booktabs}
\usepackage{algorithm}
\usepackage[noend]{algpseudocode}
\usepackage{verbatim}
\usepackage{amsmath}
\usepackage{amsthm}
\usepackage{mathabx}
\usepackage[disable]{todonotes}
\usepackage{placeins}
\usepackage{subfigure}
\usepackage{graphicx}
\usepackage{pgfplots}
\pgfplotsset{compat=1.15}
\usetikzlibrary{patterns,shapes,positioning,calc}

\usepackage{natbib}
\usepackage{geometry}
\usepackage{fontawesome5}

\pgfplotsset{pyCubeQ1Style/.style={mark=*,color=bblue}} 
\pgfplotsset{pyCubeQ2Style/.style={mark=x,color=bblue}} 
\pgfplotsset{pyCubeQ3Style/.style={mark=triangle*,color=bblue}} 
\pgfplotsset{pyCubeQ4Style/.style={mark=square*,color=bblue}} 

\pgfplotsset{
    legend image code/.code = {
        \draw[mark repeat = 2, mark phase = 2]
        plot coordinates {
            (0cm, 0cm)
            (0.15cm, 0cm)
            (0.3cm, 0cm)};}}

\pgfplotsset{JFFQ1Style/.style={mark=*,color=rred}} 
\pgfplotsset{JFFQ2Style/.style={mark=x,color=rred}} 
\pgfplotsset{JFFQ3Style/.style={mark=triangle*,color=rred}} 
\pgfplotsset{JFFQ4Style/.style={mark=square*,color=rred}} 

\pgfplotsset{JDFQ1Style/.style={mark=*,color=ggreen}} 
\pgfplotsset{JDFQ2Style/.style={mark=x,color=ggreen}} 
\pgfplotsset{JDFQ3Style/.style={mark=triangle*,color=ggreen}} 
\pgfplotsset{JDFQ4Style/.style={mark=square*,color=ggreen}} 

\pgfplotsset{SQLJQ1Style/.style={mark=*,color=ppurple}} 
\pgfplotsset{SQLJQ2Style/.style={mark=x,color=ppurple}} 
\pgfplotsset{SQLJQ3Style/.style={mark=triangle*,color=ppurple}} 
\pgfplotsset{SQLJQ4Style/.style={mark=square*,color=ppurple}} 

\definecolor{bblue}{HTML}{4f81bd}
\definecolor{bbblue}{HTML}{a5bedd}
\definecolor{rred}{HTML}{c0504d}
\definecolor{rrred}{HTML}{dfa5a4}
\definecolor{ggreen}{HTML}{9bbb59}
\definecolor{gggreen}{HTML}{c5d89f}
\definecolor{ppurple}{HTML}{9f4c7c}
\definecolor{pppurple}{HTML}{d9afc7}

\AtBeginDocument{ 
  
}

\lstdefinestyle{Python} {
  language=Python,
  morestring=[s]{"""}{"""},
  emph={access,and,break,class,continue,def,del,elif,else,
    except,exec,finally,for,from,global,if,import,in,is,
    lambda,not,or,pass,print,raise,return,try,while,
    True,False,None,self,from,import,as},
  keywordstyle=\color{black}, 
}

\newcommand{\inlinecode}[1]{\lstinline{#1}}
\newcommand{\frameworkname}{pyCube}

\newcommand{\sql}{\texttt{SQL}}

\newcommand{\pemph}[1]{{\upshape\scshape #1}}

\newcommand{\demph}[1]{\upshape\textsc{#1}}

\newcommand{\cubeinstance}{\ensuremath{c} }

\newcommand{\datacubeview}{\ensuremath{(AX, MS_v, p, \cubeinstance)} }

\begin{document}
\copyrightyear{2023}
\copyrightclause{Copyright for this paper by its authors.
  Use permitted under Creative Commons License Attribution 4.0
  International (CC BY 4.0).}
 
\conference{26th International Workshop on Design, Optimization, Languages and Analytical Processing of Big Data}
\title{Creating and Querying Data Cubes in Python using pyCube}
\author[1]{Sigmundur Vang}[%
    email=siva@cs.aau.dk,
]
\cormark[1]
\address[1]{Department of Computer Science, Aalborg University, Selma Lagerløfs Vej 300, 9220 Aalborg, Denmark}

\author[1]{Christian Thomsen}[%
    email=chr@cs.aau.dk
]

\author[1]{Torben Bach Pedersen}[%
    email=tbp@cs.aau.dk
]

\cortext[1]{Corresponding author.}

\begin{abstract}
    Data cubes are used for analyzing large data sets usually contained in data warehouses.
    The most popular data cube tools use graphical user interfaces (GUI) to do the data analysis.
    Traditionally this was fine since data analysts were not expected to be technical people.
    However, in the subsequent decades the data landscape changed dramatically requiring companies to employ large teams of highly technical data scientists in order to manage and use the ever increasing amount of data.
    These data scientists generally use tools like Python, interactive notebooks, pandas, etc. while modern data cube tools are still GUI based.
    This paper proposes a Python-based data cube tool called \frameworkname.
    \frameworkname\ is able to semi-automatically create data cubes for data stored in an RDBMS and manages the data cube metadata.
    \frameworkname's programmatic interface enables data scientist to query data cubes by specifying the expected metadata of the result.
    \frameworkname\ is experimentally evaluated on Star Schema Benchmark (SSB).
    The results show that \frameworkname\ vastly outperforms different implementations of SSB queries in pandas in both runtime and memory while being easier to read and write.
\end{abstract}

\maketitle
\todo[inline]{Readd at signs to author email}

\section{Introduction}
Data cubes are a well known and widely used logical data model and has been so ever since many companies started to use business intelligence to support business decisions.
Most data cube tools have been designed to use GUIs as their primary way of interacting with the user.
This made sense because data analysis using data cubes has traditionally been carried out by analytically intelligent but non-technical data analysts.
Since that time the technical and data landscapes have changed dramatically.
The amount of data has increased exponentially and companies save as much data as they can.
This increases demand for technical people with a new set of necessary skills and has spawned numerous different data employment roles, one of which is the data scientist.
The data scientist is expected to be more technically literate and also be able to work closely with data.
Python, pandas, spark etc. are tools commonly used by data scientists and are generally used in an interactive notebook.
Despite the change in employee skills, data cube tools are still GUI based.
While GUI based data cube tools still provide value, a programmatically based tool will better suit data scientists.
There exists ways to programmatically access data cubes one of which is MDX~\cite{whitehorn06} whose syntax and semantics inspired \frameworkname.
Unlike MDX there is not a heavy emphasis on sets and tuples in \frameworkname.
Instead ordered lists are used since \frameworkname\ constructs cube views by specifying the dimension values instead of selecting cells from the intersection of dimensions as is done in MDX.
MDX is fully declarative.
\frameworkname\ constructs cube view declaratively in a procedural environment which is an approach similar to Spark SQL~\cite{armbrust15}.
Whereas Spark SQL can be seen as \sql\ with variables, \frameworkname\ can be seen as MDX with variables.
Cube Dev's \emph{Cube}~\cite{cubedev} and Databrewery's \emph{cubes}~\cite{cubes} abstract the specific storage solutions by providing a semantic layer over the underlying data.
This semantic layer gives organisations great liberty in choosing or creating frontends that best suits their data analysis needs.
The Python library \emph{atoti}~\cite{atoti} has an in-memory data cube model and a variety of import methods making it easy to be agnostic about data storage specifics.
However, \emph{atoti} primarily interacts with data cubes through a built-in GUI dashboard which is launched directly from a Jupyter Notebook~\cite{jupyterNotebooks}.
This paper presents \frameworkname, a Python-based data cube tool for data scientists.
The tool allows data scientists to semi-automatically create data cubes from data in RDBMS.
The metadata of the data cubes is managed by \frameworkname.
The data cubes can be queried using \frameworkname's declarative API.
The result is a pandas~\cite{pandas} dataframe.
The remainder of the paper is organized as follows.
Formal definitions of terminologies are given in Section~\ref{sec:preliminaries}.
\frameworkname\ declarative API is discussed in Section~\ref{sec:use_case}.
Section~\ref{sec:populating_view} explains how \frameworkname\ computes the result.
An experimental evaluation is given in Section~\ref{sec:experiments} and a conclusion is given in Section~\ref{sec:conclusion}.

\section{Preliminaries}\label{sec:preliminaries}
This section gives mathematical definitions of all relevant concepts.
Some definitions are inspired by~\cite{ciferri13}.

A \demph{level schema} $ls = (k, A)$ is a tuple, where $k\subseteq A$ is the \demph{level key} consisting of one or more attributes, and $A$ is a set of attributes.
A \demph{level instance} of $ls$ is a function that maps from values of $k$ to tuples in $dom(a_1) \times \cdots \times dom(a_n)$
for $a_1, \ldots, a_n \in A$ where $dom(a_i)$ denotes the domain of $a_i$, i.e., the set of possible values for $a_i$.
The level instance of $ALL_{ds} = (all, \{all\})$ is $ALL_{ds}(all) = \{all\}$, where $all$ is the special \demph{ALL} attribute which denotes the top most attribute.
A \demph{dimension schema} $ds = (LS, \preceq)$ is a tuple, where $LS$ is a set of level schemas including the level schema $ALL_{ds} = (\{all\}, \{all\})$ and $\preceq$ is a partial order on the level schemas $ls\in LS$.
The $ALL_{ds}$ level schema of $df$ is the top level schema, such that $ls\preceq ALL_{ds}$ for all level schemas $ls\in LS$.
Let $LI$ be an ordered set of level instances with exactly one level instance for each level schema in LS.
A \demph{dimension instance} $d = (LM, R)$ of $ds$ is a tuple, where $LM = \{LM_1, LM_2,\dots, LM_{|LI|}\}$ is an ordered set of disjoint multisets of attribute values for each level instance $li\in LI$ called \demph{level members}, and $R = \{R_1,R_2,\dots R_{|LS|}\}$ is an ordered set of sets of \demph{roll-up} functions, one set for each set of level members, such that if $r(lm_i) = lm_j$ for some $r\in R_i$ then $ls_i\preceq ls_j$, for every $lm_i\in LM_i$ and $lm_j\in LM_j$, together with their associated level schemas $ls_i$ and $ls_j$, where $1\leq i, j\leq |LS|$.
Let $LM_\ell$ denote the set of level members associated with level instance $\ell\in LI$.

A \demph{measure schema} $ms = (a, \textsc{agg})$ is a tuple, where $a$ is an attribute with values in $\mathbb{R}$ and $\textsc{agg}$ is an aggregate function which takes in a multiset of real numbers and returns a value $r\in\mathbb{R}$.
Measure schemas can be combined with numeric operators to create calculated measure schemas.
Given two measure schemas $ms_1 = (a_1, \textsc{agg})$ and $ms_2 = (a_2, \textsc{agg})$, the expression $ms_1 * ms_2$ results in a new measure schema $(a_1 * a_2, \textsc{agg})$.
Note that the numeric operator is bound to the attribute values and not the attributes themselves and that the aggregation function in $ms_1$ and $ms_2$ must be the same.

A \demph{data cube schema} $cs = (DS, MS)$ is a tuple, where $DS$ is a set of dimension schemas, and $MS$ is a set of measure schemas.
Let $D_I$ be a set of dimension instances with exactly one dimension instance for each dimension schema in $DS$ and $LM_i$ be the set of level members in the i\textsuperscript{th} dimension instance.
Furthermore let $B_{D_I} = \bigtimes_{j = 1}^{|D_I|}LM_{j}$ be the cartesian product of the level members across all dimension instances, and let $B_{MS} = (\mathbb{R}\cup\{\bot\})^{|MS|}$ be the numerical values of the measures where $\bot$ denotes no value.
A \demph{data cube instance} $\cubeinstance : B_{D_I}\rightarrow B_{MS}$ is a function that maps combinations of level members from $B_{D_I}$ to the numerical values of the measures from $B_{MS}$.
An \demph{axis} $ax = (d, \ell, a, lm)$ is a 4-tuple, where $d$ is a dimension instance, $\ell = (k, A)\in LS$ is a level schema, $a\in A$ is an attribute on $\ell$ and $lm$ is a list of level members from $LM_\ell$.
An expression of the form  \demph{true}, \demph{false}, or $a$ $\omega$ $lit$ is a predicate where $a$ is an attribute, $\omega \in \{<, \leq, =, \neq, \geq, >\}$ is a numeric operator and $lit$ a literal.
If $p_1,p_2$ are predicates, then $(p_1)$, $p_1 \wedge p_2$, and $p_1 \vee p_2$ are predicates. The evaluation order is defined by the standard operator bindings.
A data cube view $v = (AX, MS_v, p, c)$ is a 4-tuple where $AX = \{ax_0, ax_1,\dots,ax_n\}$ and $MS_v = \{ms_1, ms_2,\dots, ms_m\}$ are ordered sets of axes and measure schemas respectively, $p$ is a predicate and \cubeinstance is the corresponding data cube instance.
The axes in a data cube view contain information on how the data in the data cube instance \cubeinstance should be presented while the measure schemas contain information on what measures should be shown.
The first measure schema in $MS_v$ is the default measure.
The predicates are applied over axes.
Data cube views are the main construct in \frameworkname.
In a \frameworkname\ session there are usually many data cube views which are constructed from querying data cube instances.
In contrast data cube instances are defined once at the beginning of a \frameworkname\ session and never change afterwards.
Each data cube instance \cubeinstance has a default data cube view $v =$ \datacubeview where $AX$ contains a 4-tuple $(d, \ell, a, lm)$ for each dimension instance $d$ in \cubeinstance where $\ell$ is the lowest level schema in the dimension instance, $a$ is the attribute that contains the set $LM_\ell$ and $lm$ is a list containing all level members from $LM_\ell$, $MS_v$ contains one measure schema from \cubeinstance and $p=$ \pemph{true}.

\section{Use case}\label{sec:use_case}
This section shows how to interact with \frameworkname\ with a scenario for a data scientist called Helle.
\subsection{Background}
Helle works for a Danish company that sells children clothes. 
The company collects data about their sales into a PostgreSQL database called salesdb.
The database is structured using the snowflake schema~\cite{kimball02} format known from the data warehouse domain.
The sales are modelled as facts, with every fact having 4 dimensions: 
a supplier, a store, a product, and a sale date. 
The measures are the total sales price and the unit sales. 
Helle gets the task of analysing the sales in 2022.
She immediately thinks that analysing the data using data cubes would be a good way to solve the task.
Helle mostly uses Python together with an interactive notebook for her data analysis tasks since it provides a single environment that enables her to both analyse the data and to compile data analysis documents.
These documents are descriptive enough to hand over to the people in need of the analysis.
She also likes to have full control over her data when she is manipulating it which is why she tries to avoid GUI tools.
However it has been difficult for Helle to incorporate data cubes in her data analysis documents using only Python due to the previously limited data cube capabilities in Python.
Therefore for the current task she intends to use the Python framework \frameworkname, together with an interactive notebook.
That way she can incorporate data cubes into her analysis while enjoying the usual benefits that she gets from using Python in an interactive notebook.

\subsection{Initializing \frameworkname}
Helle writes the expressions in Listing~\ref{lst:data_explore_example}.

\begin{lstlisting}[caption={Importing \frameworkname},label={lst:data_explore_example},numbers=left,xleftmargin=14pt,escapechar=|]
import pyCube|\label{line:data_explore1}|
postgres_engine = pyCube.engines.postgres(|\label{line:engine_start}|
    dbname="salesdb", 
    user="helle", 
    password="password123", 
    host="127.0.0.1", 
    port="5432"
    )|\label{line:engine_end}|
session = pyCube.create_session(|\label{line:session_start}|
    engine=postgres_engine)|\label{line:session_end}|
session.views|\label{line:views}|
sales_view = session.Sales|\label{line:load_view}|
sales_view.measures()|\label{line:show_measures}|
sales_view.dimensions()|\label{line:show_dimensions}|
sales_view.Date.hierarchies()|\label{line:show_hierarchy}|
sales_view.Date.year.attributes()|\label{line:show_attributes}|
sales_view.Date.year.year|\label{line:access_attribute}|
sales_view.Date.year.year[2022]|\label{line:access_attribute_value_dict}|
sales_view.Product.category.category.Blouse|\label{line:access_attribute_value_dot}|
sales_view.Date.year.year.members()|\label{line:members_example}|
sales_view.Supplier.continent.continent.Europe\|\label{line:dot_hierarchy_example_start}|
    .Denmark["Clothing Brand 1"]|\label{line:dot_hierarchy_example_end}|
sales_view.Date.year.year[2022].children()|\label{line:children_example}|
\end{lstlisting}
Helle tells \frameworkname\ to use a PostgreSQL database by creating a postgres engine with the relevant connection details in Lines~\ref{line:engine_start}-\ref{line:engine_end} and creates a \frameworkname\ session in Lines~\ref{line:session_start}-\ref{line:session_end}.
The cubes are automatically inferred by \frameworkname\ and the inference algorithm is explained in the extended version of the paper~\cite{vang23}.
The cubes inferred by \frameworkname\ are data cubes instances and default cube views are created for each cube.
Cubes are only accessed through a view in \frameworkname.
Helle lists the names of the available views in Line~\ref{line:views} which returns \inlinecode{'Sales'} and \inlinecode{'HR'}.
Helle saves the \inlinecode{Sales} view in the \inlinecode{sales_view} variable in Line~\ref{line:load_view}.
First of all Helle wants to examine the metadata of the \inlinecode{Sales} view.
Line~\ref{line:show_measures} lists the measure schemas defined in the view which returns \inlinecode{[total_sales_price, unit_sales]}.
Line~\ref{line:show_dimensions} lists the dimension schemas contained in the view which returns \inlinecode{[Supplier, Store, Product, Date]}.
Helle notices a \inlinecode{Date} dimension schema and inspects the levels of the schema with Line~\ref{line:show_hierarchy}, which returns \inlinecode{[[day, month, year, ALL]]}. 
The returned result is a list of lists of level schemas which should be interpreted as \inlinecode{day} $\rightarrow$ \inlinecode{month} $\rightarrow$ \inlinecode{year} $\rightarrow$ \inlinecode{ALL}.
The \inlinecode{Date} dimension schema only has one hierarchy so the result includes one list.
The result would include several lists if the dimension schema had multiple hierarchies.
Helle inspects the attributes on the level schema with Line~\ref{line:show_attributes} which returns \inlinecode{[year_id, year]}.
Lines~\ref{line:access_attribute} and \ref{line:access_attribute_value_dict} accesses the attributes and attribute values on the level and attribute, respectively.
Accessing attribute values whose names either contain spaces or begin with numbers uses a dictionary-like syntax, i.e. using square brackets around the value, since Python only allows variable names beginning with letters.
Line~\ref{line:access_attribute_value_dot} accesses attribute values that begin with letters using dot notation.

The \inlinecode{members} accesses multiple attribute values of the same attribute which Helle does in Line~\ref{line:members_example}.
\frameworkname\ requires the absolute path to attribute values.
Helle accesses the attribute value \inlinecode{Clothing Brand 1} in Lines~\ref{line:dot_hierarchy_example_start}-\ref{line:dot_hierarchy_example_end}.
All attribute values are in general accessed through the attribute that contains the level members in the highest level in the hierarchy.
Furthermore attribute values in a level $l$ can only be accessed from attribute values in a level directly above $l$.
The \inlinecode{children} method accesses multiple attribute values with the same parent which Helle does in Line~\ref{line:children_example}.
Additionally, the \lstinline[breaklines=false]{children} method orders the resulting attribute values.
Dates are ordered chronologically.
The default is lexicographical.

\subsection{Analyzing the data in the view}
Helle is able to group the data in the \inlinecode{sales_view} $=$ \datacubeview view using the \inlinecode{axis}, \inlinecode{measures} and \inlinecode{where} methods in \frameworkname\ as shown in Listing~\ref{lst:analyse_data_example}.
\begin{lstlisting}[caption={Include both measures in the view},label={lst:analyse_data_example},numbers=left,xleftmargin=14pt,escapechar=|]
sales_view\
  .axis(0, |\label{line:2022_start}|
        sales_view.Date
        .year.year[2022].children())\|\label{line:2022_end}|
  .axis(1, |\label{line:categories_start}|
        sales_view.Product
        .category.category.members())\|\label{line:categories_end}|
  .where(
     (sales_view.Date.day.day >= 7) |\label{line:first_week}|
     & (sales_view.Supplier.nation.nation |\label{line:danish_suppliers_start}|
        == "Denmark"))\ |\label{line:danish_suppliers_end}|
  .measures(sales_view.UnitSales)\ |\label{line:measures}|
  .output()|\label{line:view_output}|
\end{lstlisting}
The $axis(i, lm)$ method specifies the values on the axes $AX = \{(d_0, \ell_0, a_0, lm_0), (d_1, \ell_1, a_1, lm_1),\dots, (d_n, \\\ell_n, a_n, lm_n)\}$ in a view.
Recall that $AX$ is an ordered set.
The parameters are a natural number $i$, specifying what axis to modify, and a list of level members $lm$.
The dimension, level and attribute can be inferred from the level members.
The result of the \inlinecode{axis} method is a view.
Helle specifies the first axis to be the months in 2022 with Lines~\ref{line:2022_start}-\ref{line:2022_end} in Listing~\ref{lst:analyse_data_example}.
Lines~\ref{line:categories_start}-\ref{line:categories_end} specify the second axis to be the product categories.
Note that \frameworkname\ requires axis 0 to be specified before axis 1.
In general if axis $n$ is specified then axis $n-1$ should also be specified for $n>0$.
\frameworkname\ includes the \inlinecode{columns}, \lstinline[breaklines = false]{rows}, \lstinline[breaklines = false]{pages}, \inlinecode{sections} and \inlinecode{chapters} methods as aliases for the \lstinline[breaklines = false]{axis} method where the $i$ parameter have been set to $0$, $1$, $2$, $3$ and $4$, respectively.
Therefore the \inlinecode{columns(sales_view.Date.year.year[2022].children())} and \inlinecode{rows(sales_view.Product.category.category.members())} expressions are equivalent to Lines~\ref{line:2022_start}-\ref{line:2022_end} and \ref{line:categories_start}-\ref{line:categories_end}, respectively.
The $where(p)$ method, where $p$ is a predicate of the form given in the predicate definition, specifies the predicate in the data cube view.
Helle limits the view to only include the first week in every month in Line~\ref{line:first_week}.
Predicates can be combined using logical and (\inlinecode{&}) or logical or (\inlinecode{|}).
Helle furthermore limits the result to only include Danish suppliers with the \inlinecode{where} method in Lines~\ref{line:danish_suppliers_start}-\ref{line:danish_suppliers_end}.
Note that since \inlinecode{&} and \inlinecode{|} have higher precedence than the (in)equality in Python, the predicates on either side of the logical and or the logical or needs to be surrounded with parentheses.
The $measures(ms_1, ms_2,\dots)$ method specifies the values in the measure schemas $MS_v = \{ms_1, ms_2, \dots, ms_m\}$ according to the order given in the parameters.
The parameters of \inlinecode{measures} is any number of measure schemas $ms_i$.
Helle includes the unit sales measures in Line~\ref{line:measures}.
The \inlinecode{output()} method on Line~\ref{line:view_output} populates the view with values by sending an \sql\ query to the PostgreSQL database and reformatting the \sql\ result set into a pandas dataframe which is partially shown in Table~\ref{tab:outputresultexample}.
The \sql\ generation is explained in Section~\ref{sec:sql_generation}.
\begin{table}
    \caption{Some of the result from invoking \inlinecode{view.output()}}
    \begin{tabular}{@{}lcccl@{}}
        \toprule
                                     & January               & February         & March            & $\cdots$ \\
                                     \cmidrule(lr){2-2}      \cmidrule(lr){3-3} \cmidrule(lr){4-4} 
                                     & UnitSales             & UnitSales        & UnitSales        & $\cdots$ \\ 
        \midrule
        Blouse                       & 754                   & 659              & 844              & $\cdots$ \\
        Pants                        & 378                   & 129              & 928              & $\cdots$ \\
        \multicolumn{1}{c}{$\cdots$} & $\cdots$              & $\cdots$         & $\cdots$         & $\cdots$ \\ 
        \bottomrule
    \end{tabular}
    \label{tab:outputresultexample}
\end{table}
The retrieved values are obtained using the unit sales measures.
Helle is then able to analyse the sales in 2022 by identifying the most useful metadata and specifying them as parameters to the methods explained in this section.

\section{Populating the View}\label{sec:populating_view}
When populating a view $v = \datacubeview$ with values (i.e., when using the \inlinecode{output} method on $v$) an \sql\ query is generated from $v$.
The query retrieves the values of the cells in $c$.
Subsequently the retrieved values are formatted into a pandas dataframe $df$.
The different dimensions of $df$ is dictated by $v$.
Finally $df$ is returned to the user.
The \sql\ generation is described in Section~\ref{sec:sql_generation} and the formatting of the dataframe is described in Section~\ref{sec:format_dataframe}.

\subsection{Generating the \sql\ query}\label{sec:sql_generation}
Given a data cube view $v=(AX, MS_v, p, c)$, where $AX = [(d_1, \ell_1, a_1, lm_1), (d_2, \ell_2, a_2, lm_2),\dots,(d_n,\\ \ell_n, a_n, lm_n)]$, $MS_v = \{(a_1, \textsc{agg}_1), (a_2, \textsc{agg}_2),\dots,\\(a_m, \textsc{agg}_m)\}$, $p$ is a predicate and \cubeinstance is the data cube instance, then the \sql\ generated from \inlinecode{v.output()} is given in Listing~\ref{lst:generating_sql}.
The \sql\ generation is divided into three functions which are explained in the following.

The \inlinecode{get_from_clause_subset(}$d$\inlinecode{)} is a function that generates an \sql\ query subset which will denormalize the dimension tables in $d$ and join it with the fact table.
An example is shown in the extended version of the paper\cite{vang23}.
The \inlinecode{get_from_clause_subset} function is extended to take in as a parameter a list of dimensions which is equivalent to sequentially invoking the method on each individual element in the list.
The \inlinecode{inclusion_where_clause_subset(}$AX$\lstinline[breaklines = false]{)} is a function that produces a series of $\ell_i$.$a_i$ \inlinecode{IN (}$lm_i$\inlinecode{)} expressions for every $ax_i$ in $AX$.
The expressions are separated with \texttt{AND}s.
The \inlinecode{predicate_where_clause_subset(p)} is a function that produces a series of $\ell_{p_i}$.$a_{p_i}$ $\omega_i$ $lit_i$ expressions for every $p_i$ in $p$.
The expressions are separated either with an \texttt{AND} or an \texttt{OR} depending on the user-provided predicates.
Parentheses are placed according to the evaluation order.
The dimension schemas included in $p$ but not in $AX$ are indicated by $d_{p_1}, d_{p_2},\dots,d_{p_i}$.
The key on level $\ell$ is denoted by $\ell$.$k$.

\begin{lstlisting}[caption={Generated SQL query},label={lst:generating_sql},mathescape=true]
SELECT $\ell_1$.$a_1$, $\ell_2$.$a_2$, $\dots$, $\ell_n$.$a_n$, 
       AGG$_1(ft.a_1)$ AS $a_1$,
       AGG$_2(ft.a_2)$ AS $a_2$,$\dots$,
       AGG$_m(ft.a_m)$ AS $a_m$
FROM 
    ft
    get_from_clause_subset(
        [$d_1, d_2, \dots, d_n, d_{p_1}, d_{p_2},\dots,d_{p_i}$])
WHERE 
    inclusion_where_clause_subset($AX$)
    AND 
    (predicate_where_clause_subset($p$))
GROUP BY 
    $\ell_1$.$a_1$, $\ell_1$.$k$,
    $\ell_2$.$a_2$, $\ell_2$.$k$,
    $\dots$
    $\ell_n$.$a_n$, $\ell_n$.$k$
\end{lstlisting}
The generated \sql\ selects the attributes in all axes first and all measures second, both in the order they appear in the view $v$.
The measures are given the attribute name as aliases.
The fact table is joined with all dimensions used in the axes $AX$ and all tables used in the predicate $p$.
The where clause consists of the \lstinline[breaklines=false]{inclusion_where_clause_subset} and \inlinecode{predicate_where_clause_subset} methods with an \inlinecode{AND} between them.
The \inlinecode{predicate_where_clause_subset} method is surrounded by parentheses to ensure the correct order of operations.
Finally the attribute and the key of the levels in the axes $AX$ are used in the group by clause.
An example of a \frameworkname\ expression and the result set produced by the generated query are shown in Listing~\ref{lst:pyCube_query_example} and Table~\ref{tab:sql_result_example}, respectively.

\begin{lstlisting}[caption={Example \frameworkname\ expression},label={lst:pyCube_query_example}]
sales_view\
    .columns(sales_view.Date
             .year.year[2022].children())\
    .rows([
            sales_view.Product
                .category.category.Blouse,
            sales_view.Product
                .category.category.Pants
          ])\
    .pages([sales_view.Store.city.city.Aalborg])\
    .where((sales_view.Date.month.month 
                == "January") 
           | (sales_view.Date.month.month 
                == "February"))\
    .measures(sales_view.TotalSalesPrice, 
              sales-view.UnitSales)\
    .output()
\end{lstlisting}
The \frameworkname\ expression shown in Listing~\ref{lst:pyCube_query_example} generates a dataframe where months of the year 2022 are on the columns, the "Blouse" and "Pants" categories are on rows and the "Aalborg" city is on pages.
The \lstinline[breaklines = false]{where} method further specifies that the columns should only be the months of January and February and the \inlinecode{measures} method specifies that two measures should be used: \inlinecode{TotalSalesprice} and \inlinecode{UnitSales}.
The \frameworkname\ expression generates a view $(AX, MS, p, c)$ where $AX = [(Date, month, month, lm_0), (Product, category, \\ category, [Blouse, Pants]), (Store, city, city,$\\$[Aalborg])]$, $MS = \{(TotalSalesPrice, SUM), \\(UnitSales, SUM)\}$, $p = (month = January\lor month = February)$ and $c$ is the data cube instance.
The result set from the generated \sql\ is shown in Table~\ref{tab:sql_result_example}.
TSP and US are abbreviations for TotalSalesPrice and UnitSales, respectively.
An elaborated example is given in the extended version of the paper~\cite{vang23}.

\begin{table}
    \caption{The intermediary result set produced from Listing~\ref{lst:pyCube_query_example}}
    \begin{tabular}{@{}ccccc@{}}
        \toprule
        Month & Category & City & TSP & US \\ 
        \midrule
        January & Blouse & Aalborg & 946513 & 754 \\
        January & Pants & Aalborg & 846598 & 378 \\
        February & Blouse & Aalborg & 468954 & 659 \\
        February & Pants & Aalborg & 120546 & 129 \\
        \bottomrule
    \end{tabular}
    \label{tab:sql_result_example}
\end{table}

\subsection{Converting Result Sets to Dataframes}\label{sec:format_dataframe}
Listing~\ref{lst:pandas_conversion} shows the conversion of a \sql\ result set into a pandas dataframe conforming to the metadata specified by a \frameworkname\ expression.

\begin{lstlisting}[caption={Converting \sql\ result set},label={lst:pandas_conversion},numbers=left,xleftmargin=14pt,escapechar=|]
df = pd.read_sql(query, conn)|\label{line:read_sql}|
final_df = df.pivot(columns=columns, |\label{line:pivot_columns}|
                    index=rows, |\label{line:pivot_index}|
                    values=measures)|\label{line:pivot_measures}|
final_df = final_df.reorder_levels(list(|\label{line:reorder_levels_start}|
        range(1, len(columns) + 1)) + [0], axis=1)|\label{line:reorder_levels_end}|
\end{lstlisting}
Three pandas methods are used: \inlinecode{read_sql}, \inlinecode{pivot} and \inlinecode{reorder_levels}~\cite{pandas}.
The \inlinecode{read_sql} method returns, when given an \sql\ query and a connection to a database, the result set formatted as a dataframe.
The connection is created using some Python database adapter.
Thus if the \lstinline[breaklines = false]{read_sql} method is given the \sql\ query generated from Listing~\ref{lst:pyCube_query_example}, the output would be a dataframe structured as Table~\ref{tab:sql_result_example}.
The \inlinecode{pivot} method returns, when given the \inlinecode{columns}, \inlinecode{index} and \inlinecode{values} parameters, a dataframe "pivoted" according to the parameters.
The \inlinecode{columns} and \inlinecode{index} must be one or more names.
The names must be valid column names from the result set generated from the \inlinecode{output} method.
If the \inlinecode{columns} or \inlinecode{index} parameters are given multiple names, then \inlinecode{pivot} creates a hierarchical structure on the columns or rows, respectively.
If \inlinecode{df} on Line~\ref{line:read_sql} contains Table~\ref{tab:sql_result_example}, then the \inlinecode{columns}, \inlinecode{index} and \inlinecode{measures} parameters on Lines~\ref{line:pivot_columns},\ref{line:pivot_index} and \ref{line:pivot_measures} contain $[Month, City]$, $[Category]$ and $[TSP, US]$, respectively.
Finally the \inlinecode{reorder_levels} method in Lines~\ref{line:reorder_levels_start}-\ref{line:reorder_levels_end} makes sure that the measures are the furthest down in the column hierarchy.
If the generated query produced the result set shown in Table~\ref{tab:sql_result_example} then Listing~\ref{lst:pandas_conversion} produces the dataframe shown in Table~\ref{tab:dataframe_result_example}.

\begin{table}
    \caption{Dataframe produced by Listing~\ref{lst:pyCube_query_example}}
    \begin{tabular}{@{}lcccc@{}}
        \toprule
        & \multicolumn{4}{c}{Aalborg} \\
        \cmidrule(lr){2-5}
        & January & January & February & February \\ 
        \cmidrule(lr){2-2}\cmidrule(lr){3-3}\cmidrule(lr){4-4}\cmidrule(lr){5-5}
        & TSP & US & TSP & US \\ 
        \midrule
        Blouse & 946513 & 754 & 468954 & 659 \\
        Pants & 846598 & 378 & 120546 & 129 \\
        \bottomrule
    \end{tabular}
    \label{tab:dataframe_result_example}
\end{table}

\section{Experiments}\label{sec:experiments}
This section compares \frameworkname\ with three baselines based on pandas~\cite{pandas}.
The baselines are a (1) JoinFactsFirst (JFF) baseline, (2) JoinDimensionsFirst (JDF) baseline and (3) \sql Join (\sql J) baseline.
The baselines only differ in the order in which tables in the hierachies are joined.

\subsection{Experimental Setup}
JoinFactsFirst and JoinDimensionsFirst load all relevant tables into memory one by one as pandas dataframes.
JoinDimensionsFirst denormalizes all hierarchies by joining the relevant dimension tables before joining the hierarchies with the fact table.
JoinFactsFirst joins the fact table and the lowest level first.
Then the second lowest level is joined with the intermediate join result.
This process is repeated for all relevant levels in a hierarchy and for all relevant hierarchies.
Note that the joins are computed one at a time using the \lstinline[breaklines=false]{merge} method on dataframes.
Every intermediate result is stored as a new dataframe.
\sql Join formulates all joins as a single \sql\ query and stores the result as a dataframe.
After joining the tables, all baselines filter the resulting dataframes in the same manner.
Some experiments include calculated measures which are computed using the \inlinecode{apply} method on dataframes.
Finally the dataframes are reshaped into the final result using the \inlinecode{pivot_table} method on dataframes.
The baselines represent different levels of skill required by the data scientist with JoinFactsFirst being the easiest and \sql Join the hardest.
The joins, calculated measures, filtering and reshaping of the dataframes are done in the way the pandas user guide recommends~\cite{pandasUserGuide}.

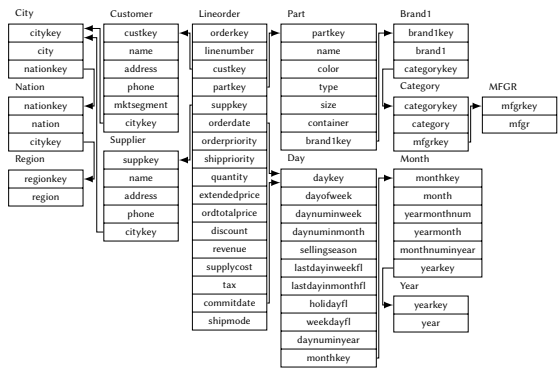
\begin{figure}
    \centering
    \resizebox{0.5\textwidth}{!}{
\begin{tikzpicture}[relation/.style = {rectangle split, rectangle split parts = #1, draw, inner ysep = 1pt, font = \tiny, minimum width = 32pt, execute at begin node = \strut}, title/.style = {font = \tiny}]
    \node[relation = 17] (lineorder) {
        orderkey
        \nodepart{two} linenumber
        \nodepart{three} custkey
        \nodepart{four} partkey
        \nodepart{five} suppkey
        \nodepart{six} orderdate
        \nodepart{seven} orderpriority
        \nodepart{eight} shippriority
        \nodepart{nine} quantity
        \nodepart{ten} extendedprice
        \nodepart{eleven} ordtotalprice
        \nodepart{twelve} discount
        \nodepart{thirteen} revenue
        \nodepart{fourteen} supplycost
        \nodepart{fifteen} tax
        \nodepart{sixteen} commitdate
        \nodepart{seventeen} shipmode
    };
    \node[title, above = 0.15cm of lineorder.north west, anchor = west] (lineordertitle) {Lineorder};

    \node[relation = 6, left = 0.2cm of lineorder.north west, anchor = north east] (customer) {
        custkey
        \nodepart{two} name
        \nodepart{three} address
        \nodepart{four} phone
        \nodepart{five} mktsegment
        \nodepart{six} citykey
    };
    \node[title, above = 0.15cm of customer.north west, anchor = west] (customertitle) {Customer};

    \node[relation = 5, below = 0.275cm of customer.south] (supplier) {
        suppkey
        \nodepart{two} name
        \nodepart{three} address
        \nodepart{four} phone
        \nodepart{five} citykey
    };
    \node[title, above = 0.15cm of supplier.north west, anchor = west] (suppliertitle) {Supplier};

    \node[relation = 3, left = 0.3cm of customer.north west, anchor = north east] (city) {
        citykey
        \nodepart{two} city
        \nodepart{three} nationkey
    };
    \node[title, above = 0.15cm of city.north west, anchor = west] (citytitle) {City};

    \node[relation = 3, below = 0.275cm of city.south] (nation) {
        nationkey
        \nodepart{two} nation
        \nodepart{three} citykey
    };
    \node[title, above = 0.15cm of nation.north west, anchor = west] (nationtitle) {Nation};

    \node[relation = 2, below = 0.275cm of nation.south] (region) {
        regionkey
        \nodepart{two} region
    };
    \node[title, above = 0.15cm of region.north west, anchor = west] (regiontitle) {Region};

    \node[relation = 7, minimum width = 41pt, right = 0.2cm of lineorder.north east, anchor = north west] (part) {
        partkey
        \nodepart{two} name
        \nodepart{three} color
        \nodepart{four} type
        \nodepart{five} size
        \nodepart{six} container
        \nodepart{seven} brand1key
    };
    \node[title, above = 0.15cm of part.north west, anchor = west] (parttitle) {Part};

    \node[relation = 3, right = 0.25cm of part.north east, anchor = north west] (brand1) {
        brand1key
        \nodepart{two} brand1
        \nodepart{three} categorykey
    };
    \node[title, above = 0.15cm of brand1.north west, anchor = west] (brand1title) {Brand1};

    \node[relation = 3, below = 0.275cm of brand1.south] (category) {
        categorykey
        \nodepart{two} category
        \nodepart{three} mfgrkey
    };
    \node[title, above = 0.15cm of category.north west, anchor = west] (categorytitle) {Category};

    \node[relation = 2, right = 0.2cm of category.north east, anchor = north west] (mfgr) {
        mfgrkey
        \nodepart{two} mfgr
    };
    \node[title, above = 0.15cm of mfgr.north west, anchor = west] (mfgrtitle) {MFGR};

    \node[relation = 11, minimum width = 41pt, below = 0.275cm of part.south] (day) {
        daykey
        \nodepart{two} dayofweek
        \nodepart{three} daynuminweek
        \nodepart{four} daynuminmonth
        \nodepart{five} sellingseason
        \nodepart{six} lastdayinweekfl
        \nodepart{seven} lastdayinmonthfl
        \nodepart{eight} holidayfl
        \nodepart{nine} weekdayfl
        \nodepart{ten} daynuminyear
        \nodepart{eleven} monthkey
    };
    \node[title, above = 0.15cm of day.north west, anchor = west] (daytitle) {Day};

    \node[relation = 6, right = 0.25cm of day.north east, anchor = north west] (month) {
        monthkey
        \nodepart{two} month
        \nodepart{three} yearmonthnum
        \nodepart{four} yearmonth
        \nodepart{five} monthnuminyear
        \nodepart{six} yearkey
    };
    \node[title, above = 0.15cm of month.north west, anchor = west] (monthtitle) {Month};

    \node[relation = 2, below = 0.275cm of month.south west, anchor = north west] (year) {
        yearkey
        \nodepart{two} year
    };
    \node[title, above = 0.15cm of year.north west, anchor = west] (yeartitle) {Year};

    \draw[-latex] (lineorder.three west) -- ++(-0.03,0) |- (customer.one east);
    \draw[-latex] (lineorder.five west) -- ++(-0.03,0) |- (supplier.one east);

    \draw[-latex] (customer.six west) -- ++(-0.05,0) |- ($(city.one east) + (0,0.07)$);
    \draw[-latex] (supplier.five west) -- ++(-0.1,0) |- ($(city.one east) + (0,-0.07)$);
    \draw[-latex] (city.three east) -- ++(0.16,0) |- (nation.one east);
    \draw[-latex] (nation.three east) -- ++(0.16,0) |- (region.one east);

    \draw[-latex] (lineorder.four east) -- ++(0.03,0) |- (part.one west);
    \draw[-latex] (part.seven east) -- ++(0.03,0) |- (brand1.one west);
    \draw[-latex] (brand1.three west) -- ++(-0.16,0) |- (category.one west);
    \draw[-latex] (category.three east) -- ++(0.03,0) |- (mfgr.one west);

    \draw[-latex] (lineorder.six east) -- ++(0.03,0) |- ($(day.one west) + (0,0.07)$);
    \draw[-latex] (lineorder.sixteen east) -- ++(0.03,0) |- ($(day.one west) + (0,-0.07)$);
    \draw[-latex] (day.eleven east) -- ++(0.03,0) |- (month.one west);
    \draw[-latex] (month.six west) -- ++(-0.16,0) |- (year.one west);
\end{tikzpicture}
}
    \caption{The snowflake schema for SSB}
    \label{fig:ssbsnowflake}
\end{figure}

The database schema (shown in Figure~\ref{fig:ssbsnowflake}) used in the experiments is a snowflaked Star Schema Benchmark (SSB)~\cite{ssb}.
Data is generated from SSB's data generator and converted into the correct schema.
The fact table has 6,000,000 rows multiplied by a scale factor.
The total data size used in the experiments is 581MB for a scale factor of 1.
The SSB is based on the popular TPC-H benchmark.
The queries in SSB are grouped into four groups named query flights.
All queries from SSB are used and are implemented four times: once using \frameworkname\ and once for each of the baselines using pandas.
All query and implementation combinations are executed five times in randomized order and the highest and lowest values are discarded.
The final result is the average of the remaining values.
Unless otherwise noted, all experiments are run on a Arch linux machine with kernel version 6.4.12-arch1-1, four cores of 11th Gen Intel Core i5-1135G7 running at a 2.40GHz clock frequency and 32GB RAM using PostgreSQL 15.4, Python 3.11 and pandas 2.0.1.
The system is similar to what a data scientist would use.

\subsection{Data Retrieval Speeds}
\begin{figure*}
    \subfigure{\begin{tikzpicture}
    \begin{axis}[
        ymajorgrids = true,
        ytick={0,25,50,75,100,125},
        xtick={0,1,2,5,10},
        width = 0.26*\linewidth,
        height = 4cm,
        xlabel = {Scale Factor},
        ylabel = {Runtime (s)},
        ymin = 0, 
        ymax = 140,
        title = {QF 1},
        legend = false,
        ]
        \addplot[pyCubeQ1Style]
            plot coordinates {(1, 0.223) (2, 0.396) (5, 1.284) (10, 2.448)};
            \addlegendentry{pycube\_q1}
        \addplot[pyCubeQ2Style]
            plot coordinates {(1, 0.205) (2, 0.366) (5, 1.269) (10, 2.466)};
            \addlegendentry{pycube\_q2}
        \addplot[pyCubeQ3Style]
            plot coordinates {(1, 0.216) (2, 0.380) (5, 1.252) (10, 2.281)};
            \addlegendentry{pycube\_q3}
        \addplot[JFFQ1Style]
            plot coordinates {(1, 29.991) (2, 58.135)};
            \addlegendentry{JFF\_q1}
        \node[anchor = south,xshift=4pt,yshift=2pt] (JFFOoM) at (axis cs:2, 58.135) {\tiny Out of Memory};
        \addplot[JFFQ2Style]
            plot coordinates {(1, 27.991) (2, 53.798)};
            \addlegendentry{JFF\_q2}
        \addplot[JFFQ3Style]
            plot coordinates {(1, 29.674) (2, 57.301)};
            \addlegendentry{JFF\_q3}
        \addplot[JDFQ1Style]
            plot coordinates {(1, 10.154) (2, 20.641) (5, 61.472) (10, 108.080)};
            \addlegendentry{JDF\_q1}
        \addplot[JDFQ2Style]
            plot coordinates {(1, 9.733) (2, 19.722) (5, 57.930) (10, 112.361)};
            \addlegendentry{JDF\_q2}
        \addplot[JDFQ3Style]
            plot coordinates {(1, 10.249) (2, 20.957) (5, 60.090) (10, 118.270)};
            \addlegendentry{JDF\_q3}
        \addplot[SQLJQ1Style]
            plot coordinates {(1, 11.762) (2, 21.665) (5, 55.183) (10, 121.609)};
            \addlegendentry{\sql J\_q1}
        \addplot[SQLJQ2Style]
            plot coordinates {(1, 9.932) (2, 19.768) (5, 53.304) (10, 112.151)};
            \addlegendentry{\sql J\_q2}
        \addplot[SQLJQ3Style]
            plot coordinates {(1, 11.918) (2, 22.223) (5, 61.449) (10, 125.813)};
            \addlegendentry{\sql J\_q3}
        \legend{}
    \end{axis}
\end{tikzpicture}}
    \subfigure{\begin{tikzpicture}
    \begin{axis}[
        ymajorgrids = true,
        ytick={0,15,30,45,60,75},
        xtick={0,1,2,5,10},
        width = 0.27*\linewidth,
        height = 4cm,
        xlabel = {Scale Factor},
        ymin = 0, 
        ymax = 90,
        title = {QF 2},
        ]

        \addplot[pyCubeQ1Style]
            plot coordinates {(1, 0.325) (2, 0.584) (5, 2.365) (10, 4.218)};
            \addlegendentry{pycube\_q1}
        \addplot[pyCubeQ2Style]
            plot coordinates {(1, 0.311) (2, 0.432) (5, 1.712) (10, 3.353)};
            \addlegendentry{pycube\_q2}
        \addplot[pyCubeQ3Style]
            plot coordinates {(1, 0.273) (2, 0.375) (5, 2.107) (10, 3.055)};
            \addlegendentry{pycube\_q3}

        \addplot[JFFQ1Style]
            plot coordinates {(1, 57.641)};
            \addlegendentry{JFF\_q1}
        \addplot[JFFQ2Style]
            plot coordinates {(1, 56.064)};
            \addlegendentry{JFF\_q2}
        \addplot[JFFQ3Style]
            plot coordinates {(1, 52.646)};
            \addlegendentry{JFF\_q3}

        \addplot[JDFQ1Style]
            plot coordinates {(1, 13.613) (2, 26.941) (5, 75.688)};
            \addlegendentry{JDF\_q1}
        \addplot[JDFQ2Style]
            plot coordinates {(1, 13.933) (2, 26.599) (5, 77.712)};
            \addlegendentry{JDF\_q2}
        \addplot[JDFQ3Style]
            plot coordinates {(1, 13.708) (2, 26.253) (5, 71.943)};
            \addlegendentry{JDF\_q3}
        \node[anchor = west, xshift = -30pt] (JDFOoM) at (axis cs:10, 77.712) {\tiny Out of Memory};

        \addplot[SQLJQ1Style]
            plot coordinates {(1, 16.354) (2, 26.303) (5, 83.828)};
            \addlegendentry{\sql J\_q1}
        \addplot[SQLJQ2Style]
            plot coordinates {(1, 13.584) (2, 24.559) (5, 74.037)};
            \addlegendentry{\sql J\_q2}
        \addplot[SQLJQ3Style]
            plot coordinates {(1, 14.156) (2, 24.199) (5, 79.027)};
            \addlegendentry{\sql J\_q3}
        \legend{}
    \end{axis}
    \node[rotate=45] (JFFOoM) at (0.40, 2.00) {\tiny Out of memory};
\end{tikzpicture}}
    \subfigure{\begin{tikzpicture}
    \begin{axis}[
        ymajorgrids = true,
        ytick={0,15,30,45,60,75},
        xtick={0,1,2,5,10},
        width = 0.27*\linewidth,
        height = 4cm,
        xlabel = {Scale Factor},
        ymin = 0, 
        ymax = 90,
        title = {QF 3},
        ]

        \addplot[pyCubeQ1Style]
            plot coordinates {(1, 0.573) (2, 1.088) (5, 3.697) (10, 4.473)};
            \addlegendentry{pycube\_q1}
        \addplot[pyCubeQ2Style]
            plot coordinates {(1, 0.336) (2, 0.507) (5, 2.363) (10, 3.097)};
            \addlegendentry{pycube\_q2}
        \addplot[pyCubeQ3Style]
            plot coordinates {(1, 0.201) (2, 0.361) (5, 2) (10, 2.738)};
            \addlegendentry{pycube\_q3}
        \addplot[pyCubeQ4Style]
            plot coordinates {(1, 0.223) (2, 0.364) (5, 1.681) (10, 2.734)};
            \addlegendentry{pycube\_q4}

        \addplot[JFFQ1Style]
            plot coordinates {(1, 17.968) (2, 35.742)};
            \addlegendentry{JFF\_q1}
        \addplot[JFFQ2Style]
            plot coordinates {(1, 16.94) (2, 32.629)};
            \addlegendentry{JFF\_q2}
        \addplot[JFFQ3Style]
            plot coordinates {(1, 15.762) (2, 29.354)};
            \addlegendentry{JFF\_q3}
        \addplot[JFFQ4Style]
            plot coordinates {(1, 15.35) (2, 30.736)};
            \addlegendentry{JFF\_q4}
        \node[anchor = south, rotate = 60, xshift=5pt] (JFFOoM) at (axis cs: 2, 35.7) {\tiny Out of Memory};

        \addplot[JDFQ1Style]
            plot coordinates {(1, 14.32) (2, 27.104)};
            \addlegendentry{JDF\_q1}
        \addplot[JDFQ2Style]
            plot coordinates {(1, 13.878) (2, 26.696) (5, 73.345)};
            \addlegendentry{JDF\_q2}
        \addplot[JDFQ3Style]
            plot coordinates {(1, 13.98) (2, 26.175) (5, 77.644)};
            \addlegendentry{JDF\_q3}
        \addplot[JDFQ4Style]
            plot coordinates {(1, 14.131) (2, 27.186) (5, 77.292)};
            \addlegendentry{JDF\_q4}
        \node[anchor = west, xshift = -30pt] (JDFOoM) at (axis cs: 10, 77.6) {\tiny Out of Memory};

        \addplot[SQLJQ1Style]
            plot coordinates {(1, 15.468) (2, 27.644) (5, 84.242)};
            \addlegendentry{\sql J\_q1}
        \addplot[SQLJQ2Style]
            plot coordinates {(1, 15.004) (2, 25.919) (5, 67.497)};
            \addlegendentry{\sql J\_q2}
        \addplot[SQLJQ3Style]
            plot coordinates {(1, 12.903) (2, 22.08 ) (5, 70.335)};
            \addlegendentry{\sql J\_q3}
        \addplot[SQLJQ4Style]
            plot coordinates {(1, 13.568) (2, 23.867) (5, 75.916)};
            \addlegendentry{\sql J\_q4}
        \legend{}
    \end{axis}
\end{tikzpicture}}
    \subfigure{\begin{tikzpicture}
    \begin{axis}[
        ymajorgrids = true,
        ytick={0,15,30,45,60,75},
        xtick={0,1,2,5,10},
        width = 0.25*\linewidth,
        height = 4cm,
        xlabel = {Scale Factor},
        ymin = 0, 
        ymax = 70,
        title = {QF 4},
        legend cell align = left,
        legend columns = 2,
        legend style = {
            at = {(1.05, 1)},
            anchor = north west,
            font = \tiny,
            mark options = {scale = 0.5},
            row sep = -0.5pt,
        },
        ]

        \addplot[pyCubeQ1Style]
            plot coordinates {(1, 0.788) (2, 1.492) (5, 5.622) (10, 7.865)};
            \addlegendentry{pyc\_q1}
        \addplot[pyCubeQ2Style]
            plot coordinates {(1, 1.023) (2, 2.004) (5, 4.308) (10, 7.387)};
            \addlegendentry{pyc\_q2}
        \addplot[pyCubeQ3Style]
            plot coordinates {(1, 0.328) (2, 0.619) (5, 2.111) (10, 3.105)};
            \addlegendentry{pyc\_q3}
        \addlegendimage{pyCubeQ4Style}
        \addlegendentry{pyc\_q4}

        \addplot[JFFQ1Style]
            plot coordinates {(1, 30.823) (2, 59.761)};
            \addlegendentry{JFF\_q1}
        \addplot[JFFQ2Style]
            plot coordinates {(1, 32.082) (2, 56.566)};
            \addlegendentry{JFF\_q2}
        \addplot[JFFQ3Style]
            plot coordinates {(1, 30.19) (2, 55.869)};
            \addlegendentry{JFF\_q3}
        \addlegendimage{JFFQ4Style}
        \addlegendentry{JFF\_q4}
        \node[xshift = 1pt] (JFFOoM) at (axis cs: 5, 57.7) {\tiny Out of Memory};

        \addplot[JDFQ1Style]
            plot coordinates {(1, 20.948) (2, 40.645)};
            \addlegendentry{JDF\_q1}
        \addplot[JDFQ2Style]
            plot coordinates {(1, 20.968) (2, 40.252)};
            \addlegendentry{JDF\_q2}
        \addplot[JDFQ3Style]
            plot coordinates {(1, 21.456) (2, 39.623)};
            \addlegendentry{JDF\_q3}
        \addlegendimage{JDFQ4Style}
        \addlegendentry{JDF\_q4}
        \node[xshift = 1pt] (JDFOoM) at (axis cs: 5, 39) {\tiny Out of Memory};

        \addplot[SQLJQ1Style]
            plot coordinates {(1, 19.935) (2, 40.308)};
            \addlegendentry{\sql J\_q1}
        \addplot[SQLJQ2Style]
            plot coordinates {(1, 20.067) (2, 38.626)};
            \addlegendentry{\sql J\_q2}
        \addplot[SQLJQ3Style]
            plot coordinates {(1, 19.686) (2, 34.943)};
            \addlegendentry{\sql J\_q3}
        \addlegendimage{SQLJQ4Style}
        \addlegendentry{\sql J\_q4}

    \end{axis}
\end{tikzpicture}}
    \caption{Laptop runtimes growth over scale factors}
    \label{fig:laptop_runtime_comparison_scale_graph}
\end{figure*}
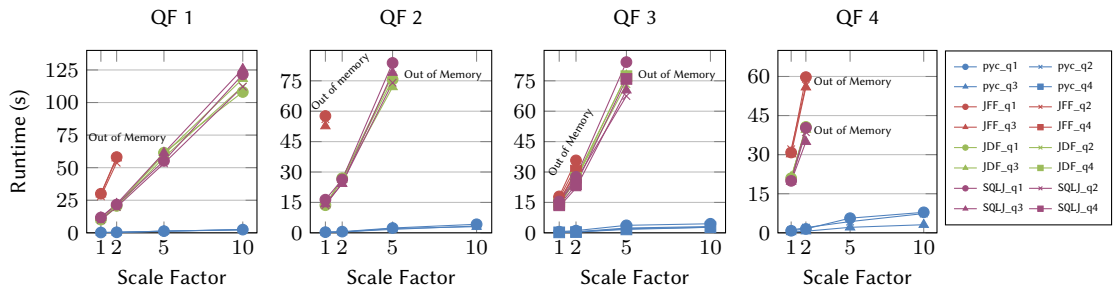

Figure~\ref{fig:laptop_runtime_comparison_scale_graph} compares the runtime performance of \frameworkname\ and the baselines for scale factors 1, 2, 5 and 10.
The runtime is in seconds of wall clock time and is measured using Python's \inlinecode{time} module.

\begin{figure*}
    \subfigure{\begin{tikzpicture}[
  every axis/.style={
    ymajorgrids = true,
    major x tick style = transparent,
    xtick = data,
    enlarge x limits=0.25,
    symbolic x coords={Q11,Q12,Q13},
    width  = 0.28*\linewidth,
    height = 4cm,
    ylabel = {Runtime (s)},
    y label style = {font=\footnotesize,at={(-0.2,0.5)}},
    ybar stacked,
    ybar=1.2pt,
    ymin=0,
    ymax=35,
    scaled y ticks = false,
    bar width=4pt,
    legend cell align=left,
    legend style={
            at={(1,1.05)},
            anchor=south east,
            column sep=1ex
    },
  },
]\label{fig:QF1SF1}

\begin{axis}[bar shift=-9pt]
    \addplot[style={color=bblue,fill=bblue}]
        coordinates {(Q11, 0.008) (Q12, 0.008) (Q13, 0.008)};
        \addlegendentry{\frameworkname\ Python}
    \addplot[style={color=bbblue,fill=bbblue}]
        coordinates {(Q11, 0.215) (Q12, 0.197) (Q13, 0.208)};
        \addlegendentry{\frameworkname\ DB}
\end{axis}

\node[below right,rotate=90,yshift=3.5pt,xshift=-4pt,style={bblue}] (Q11) at (0.05, 0.075) {0.223};
\node[below right,rotate=90,yshift=-20pt,xshift=-4pt,style={bblue}] (Q12) at (0.05, 0.075) {0.205};
\node[below right,rotate=90,yshift=-44pt,xshift=-4pt,style={bblue}] (Q13) at (0.05, 0.075) {0.216};

\begin{axis}[bar shift=-3pt,hide axis]
    \addplot+[style={color=rred,fill=rred}]
        coordinates {(Q11, 18.027) (Q12, 15.795) (Q13, 17.564)};
    \addplot+[style={color=rrred,fill=rrred}]
        coordinates {(Q11, 11.964) (Q12, 12.196) (Q13, 12.11)};
\end{axis}

\begin{axis}[bar shift=3pt,hide axis]
    \addplot+[style={color=ggreen,fill=ggreen}]
        coordinates {(Q11, 4.476) (Q12, 4.069) (Q13, 4.581)};
    \addplot+[style={color=gggreen,fill=gggreen}]
        coordinates {(Q11, 5.678) (Q12, 5.664) (Q13, 5.668)};
\end{axis}

\begin{axis}[bar shift=9pt,hide axis]
    \addplot+[style={color=ppurple,fill=ppurple}]
        coordinates {(Q11, 4.265) (Q12, 3.28) (Q13, 4.214)};
    \addplot+[style={color=pppurple,fill=pppurple}]
        coordinates {(Q11, 7.497) (Q12, 6.652) (Q13, 7.704)};
\end{axis}

\end{tikzpicture}}
    \subfigure{\begin{tikzpicture}[
  every axis/.style={
    ymajorgrids = true,
    major x tick style = transparent,
    xtick = data,
    enlarge x limits=0.25,
    symbolic x coords={
      Q21,
      Q22,
      Q23,
    },
    width  = 0.28*\textwidth,
    height = 4cm,
    y label style = {font=\footnotesize,at={(-0.05,0.5)}},
    ybar stacked,
    ybar=1.2pt,
    ymin=0,
    ymax=65,
    scaled y ticks = false,
    bar width=4pt,
    legend cell align=left,
    legend style={
            at={(1,1.05)},
            anchor=south east,
            column sep=1ex
    },
  },
]

\begin{axis}[bar shift=-9pt]
    \addplot[style={color=bblue,fill=bblue}]
        coordinates {(Q21, 0.015) (Q22, 0.016) (Q23, 0.017)};
    \addplot[style={color=bbblue,fill=bbblue}]
        coordinates {(Q21, 0.31) (Q22, 0.295) (Q23, 0.256)};
\end{axis}

\node[below right,rotate=90,yshift=3.5pt,xshift=-4pt,style={bblue}] (Q11) at (0.05, 0.075) {0.325};
\node[below right,rotate=90,yshift=-20pt,xshift=-4pt,style={bblue}] (Q12) at (0.05, 0.075) {0.311};
\node[below right,rotate=90,yshift=-44pt,xshift=-4pt,style={bblue}] (Q13) at (0.05, 0.075) {0.273};

\begin{axis}[bar shift=-3pt,hide axis]
    \addplot+[style={color=rred,fill=rred}]
        coordinates {(Q21, 45.325) (Q22, 42.952) (Q23, 41.01)};
        \addlegendentry{JFF Python}
    \addplot+[style={color=rrred,fill=rrred}]
        coordinates {(Q21, 12.316) (Q22, 13.112) (Q23, 11.636)};
        \addlegendentry{JFF DB}
\end{axis}

\begin{axis}[bar shift=3pt,hide axis]
    \addplot+[style={color=ggreen,fill=ggreen}]
        coordinates {(Q21, 7.618) (Q22, 7.614) (Q23, 7.454)};
    \addplot+[style={color=gggreen,fill=gggreen}]
        coordinates {(Q21, 5.995) (Q22, 6.319) (Q23, 6.254)};
\end{axis}

\begin{axis}[bar shift=9pt,hide axis]
    \addplot+[style={color=ppurple,fill=ppurple}]
        coordinates {(Q21, 3.283) (Q22, 2.923) (Q23, 2.863)};
    \addplot+[style={color=pppurple,fill=pppurple}]
        coordinates {(Q21, 13.071) (Q22, 10.661) (Q23, 11.293)};
\end{axis}

\end{tikzpicture}}
    \subfigure{\begin{tikzpicture}[
  every axis/.style={
    ymajorgrids = true,
    major x tick style = transparent,
    xtick = data,
    enlarge x limits=0.25,
    symbolic x coords={
      Q31,
      Q32,
      Q33,
      Q34,
    },
    width  = 0.36*\textwidth,
    height = 4cm,
    y label style = {font=\footnotesize,at={(-0.05,0.5)}},
    ybar stacked,
    ybar=1.2pt,
    ymin=0,
    ymax=20,
    scaled y ticks = false,
    bar width=4pt,
    legend cell align=left,
    legend style={
            at={(1,1.05)},
            anchor=south east,
            column sep=1ex
    },
  },
]

\begin{axis}[bar shift=-9pt]
    \addplot[style={color=bblue,fill=bblue}]
        coordinates {(Q31, 0.016) (Q32, 0.02) (Q33, 0.017) (Q34, 0.018)};
    \addplot[style={color=bbblue,fill=bbblue}]
        coordinates {(Q31, 0.557) (Q32, 0.316) (Q33, 0.184) (Q34, 0.205)};
\end{axis}

\node[below right,rotate=90,yshift=-3pt,xshift=0pt,style={bblue}] (Q31) {0.573};
\node[below right,rotate=90,yshift=-26.5pt,xshift=-1pt,style={bblue}] (Q32) {0.336};
\node[below right,rotate=90,yshift=-50pt,xshift=-1.5pt,style={bblue}] (Q33) {0.201};
\node[below right,rotate=90,yshift=-73.5pt,xshift=-1.5pt,style={bblue}] (Q34) {0.223};

\begin{axis}[bar shift=-3pt,hide axis]
    \addplot+[style={color=rred,fill=rred}]
        coordinates {(Q31, 12.102) (Q32, 10.791) (Q33, 9.319) (Q34, 9.481)};
    \addplot+[style={color=rrred,fill=rrred}]
        coordinates {(Q31, 5.866) (Q32, 6.149) (Q33, 6.443) (Q34, 5.869)};
\end{axis}

\begin{axis}[bar shift=3pt,hide axis]
    \addplot+[style={color=ggreen,fill=ggreen}]
        coordinates {(Q31, 8.159) (Q32, 7.724) (Q33, 7.691) (Q34, 7.972)};
        \addlegendentry{JDF Python}
    \addplot+[style={color=gggreen,fill=gggreen}]
        coordinates {(Q31, 6.161) (Q32, 6.154) (Q33, 6.289) (Q34, 6.159)};
        \addlegendentry{JDF DB}
\end{axis}

\begin{axis}[bar shift=9pt,hide axis]
    \addplot+[style={color=ppurple,fill=ppurple}]
        coordinates {(Q31, 3.314) (Q32, 3.293) (Q33, 3.335) (Q34, 3.697)};
    \addplot+[style={color=pppurple,fill=pppurple}]
        coordinates {(Q31, 12.154) (Q32, 11.711) (Q33, 9.568) (Q34, 9.871)};
\end{axis}

\end{tikzpicture}}
    \subfigure{\begin{tikzpicture}[
  every axis/.style={
    ymajorgrids = true,
    major x tick style = transparent,
    xtick = data,
    enlarge x limits=0.25,
    symbolic x coords={Q41,Q42,Q43},
    width  = 0.28*\textwidth,
    height = 4cm,
    ylabel={Scale factor 1},
    y label style = {font=\small,at={(1.15,0.5)}},
    ybar stacked,
    ybar=1.2pt,
    ymin=0,
    ymax=35,
    scaled y ticks = false,
    bar width=4pt,
    legend cell align=left,
    legend style={
            at={(1,1.05)},
            anchor=south east,
            column sep=1ex
    },
  },
]

\begin{axis}[bar shift=-9pt]
    \addplot[style={color=bblue,fill=bblue}]
        coordinates {(Q41, 0.014) (Q42, 0.017) (Q43, 0.017)};
    \addplot[style={color=bbblue,fill=bbblue}]
        coordinates {(Q41, 0.774) (Q42, 1.006) (Q43, 0.311)};
\end{axis}

\node[below right,rotate=90,yshift=2.5pt,xshift=-0.5pt,style={bblue}] (Q41) {0.788};
\node[below right,rotate=90,yshift=-21.5pt,xshift=-0.5pt,style={bblue}] (Q42) {1.023};
\node[below right,rotate=90,yshift=-45.5pt,xshift=-1.5pt,style={bblue}] (Q43) {0.328};

\begin{axis}[bar shift=-3pt,hide axis]
    \addplot+[style={color=rred,fill=rred}]
        coordinates {(Q41, 23.865) (Q42, 24.174) (Q43, 22.854)};
    \addplot+[style={color=rrred,fill=rrred}]
        coordinates {(Q41, 6.958) (Q42, 7.908) (Q43, 7.336)};
\end{axis}

\begin{axis}[bar shift=3pt,hide axis]
    \addplot+[style={color=ggreen,fill=ggreen}]
        coordinates {(Q41, 13.958) (Q42, 13.702) (Q43, 13.801)};
    \addplot+[style={color=gggreen,fill=gggreen}]
        coordinates {(Q41, 6.99) (Q42, 7.266) (Q43, 7.655)};
\end{axis}

\begin{axis}[bar shift=9pt,hide axis]
    \addplot+[style={color=ppurple,fill=ppurple}]
        coordinates {(Q41, 6.053) (Q42, 4.746) (Q43, 4.152)};
        \addlegendentry{\sql J Python}
    \addplot+[style={color=pppurple,fill=pppurple}]
        coordinates {(Q41, 13.882) (Q42, 15.321) (Q43, 15.534)};
        \addlegendentry{\sql J DB}
\end{axis}

\end{tikzpicture}}
    \subfigure{\begin{tikzpicture}[
  every axis/.style={
    ymajorgrids = true,
    major x tick style = transparent,
    xtick = data,
    enlarge x limits=0.25,
    symbolic x coords={
      Q11,
      Q12,
      Q13,
    },
    width  = 0.28*\textwidth,
    height = 4cm,
    ylabel = {Runtime (s)},
    y label style = {font=\footnotesize,at={(-0.2,0.5)}},
    ybar stacked,
    ybar=1.2pt,
    ymin=0,
    ymax=130,
    scaled y ticks = false,
    bar width=4pt,
    legend cell align=left,
    legend style={
            at={(1,1.05)},
            anchor=south east,
            column sep=1ex
    },
  },
]
\begin{axis}[bar shift=-9pt]
    \addplot[style={color=bblue,fill=bblue}]
        coordinates {(Q11, 0.012) (Q12, 0.012) (Q13, 0.011)};
    \addplot[style={color=bbblue,fill=bbblue}]
        coordinates {(Q11, 2.436) (Q12, 2.454) (Q13, 2.27)};
\end{axis}

\node[below right,rotate=90,yshift=3.1pt,xshift=-1pt,style={bblue}] (Q11) at (0.03, 0.01) {2.448};
\node[below right,rotate=90,yshift=-20.5pt,xshift=-1pt,style={bblue}] (Q12) at (0.03, 0.01) {2.466};
\node[below right,rotate=90,yshift=-44.5pt,xshift=-1pt,style={bblue}] (Q13) at (0.03, 0.01) {2.281};

\node[below right,rotate=90,yshift=-5pt,xshift=-2.5pt,style={rred}] (Q21) at (0, 0.01)  {\tiny Out of Memory};
\node[below right,rotate=90,yshift=-29pt,xshift=-2.5pt,style={rred}] (Q22) at (0, 0.01) {\tiny Out of Memory};
\node[below right,rotate=90,yshift=-53pt,xshift=-2.5pt,style={rred}] (Q23) at (0, 0.01) {\tiny Out of Memory};

\begin{axis}[bar shift=3pt,hide axis]
    \addplot+[style={color=ggreen,fill=ggreen}]
        coordinates {(Q11, 45.308) (Q12, 43.977) (Q13, 49.931)};
    \addplot+[style={color=gggreen,fill=gggreen}]
        coordinates {(Q11, 62.772) (Q12, 68.384) (Q13, 68.339)};
\end{axis}
\begin{axis}[bar shift=9pt,hide axis]
    \addplot+[style={color=ppurple,fill=ppurple}]
        coordinates {(Q11, 43.377) (Q12, 36.923) (Q13, 43.382)};
    \addplot+[style={color=pppurple,fill=pppurple}]
        coordinates {(Q11, 78.232) (Q12, 75.228) (Q13, 82.431)};
\end{axis}
\end{tikzpicture}}
    \subfigure{\begin{tikzpicture}[
  every axis/.style={
    ymajorgrids = true,
    major x tick style = transparent,
    xtick = data,
    enlarge x limits=0.25,
    symbolic x coords={
      Q21,
      Q22,
      Q23,
    },
    width  = 0.28*\textwidth,
    height = 4cm,
    y label style = {font=\footnotesize,at={(-0.05,0.5)}},
    ybar stacked,
    ybar=1.2pt,
    ymin=0,
    ymax=5,
    scaled y ticks = false,
    bar width=4pt,
    legend cell align=left,
    legend style={
            at={(1,1.05)},
            anchor=south east,
            column sep=1ex
    },
  },
]
\begin{axis}[bar shift=-9pt]
    \addplot[style={color=bblue,fill=bblue}]
        coordinates {(Q21, 0.021) (Q22, 0.02) (Q23, 0.02)};
    \addplot[style={color=bbblue,fill=bbblue}]
        coordinates {(Q21, 4.197) (Q22, 3.333) (Q23, 3.035)};
\end{axis}

\node[below right,rotate=90,yshift=-4pt,xshift=-2.5pt,style={rred}] (Q11) at (0, 0.01) {\tiny Out of Memory};
\node[below right,rotate=90,yshift=-28pt,xshift=-2.5pt,style={rred}] (Q12) at (0, 0.01) {\tiny Out of Memory};
\node[below right,rotate=90,yshift=-52pt,xshift=-2.5pt,style={rred}] (Q13) at (0, 0.01) {\tiny Out of Memory};

\node[below right,rotate=90,yshift=-9pt,xshift=-2.5pt,style={ggreen}] (Q11a) at (0, 0.01) {\tiny Out of Memory};
\node[below right,rotate=90,yshift=-33pt,xshift=-2.5pt,style={ggreen}] (Q12a) at (0, 0.01) {\tiny Out of Memory};
\node[below right,rotate=90,yshift=-57pt,xshift=-2.5pt,style={ggreen}] (Q13a) at (0, 0.01) {\tiny Out of Memory};

\node[below right,rotate=90,yshift=-14pt,xshift=-2.5pt,style={ppurple}] (Q11b) at (0, 0.01) {\tiny Out of Memory};
\node[below right,rotate=90,yshift=-38pt,xshift=-2.5pt,style={ppurple}] (Q12b) at (0, 0.01) {\tiny Out of Memory};
\node[below right,rotate=90,yshift=-62pt,xshift=-2.5pt,style={ppurple}] (Q13b) at (0, 0.01) {\tiny Out of Memory};

\end{tikzpicture}}
    \subfigure{\begin{tikzpicture}[
  every axis/.style={
    ymajorgrids = true,
    major x tick style = transparent,
    xtick = data,
    enlarge x limits=0.25,
    symbolic x coords={
      Q31,
      Q32,
      Q33,
      Q34,
    },
    width  = 0.36*\textwidth,
    height = 4cm,
    y label style = {font=\footnotesize,at={(-0.05,0.5)}},
    ybar stacked,
    ybar=1.2pt,
    ymin=0,
    ymax=5,
    scaled y ticks = false,
    bar width=4pt,
    legend cell align=left,
    legend style={
            at={(1,1.05)},
            anchor=south east,
            column sep=1ex
    },
  },
]
\begin{axis}[bar shift=-9pt]
    \addplot[style={color=bblue,fill=bblue}]
        coordinates {(Q31, 0.019) (Q32, 0.023) (Q33, 0.023) (Q34, 0.023)};
    \addplot[style={color=bbblue,fill=bbblue}]
        coordinates {(Q31, 4.454) (Q32, 3.074) (Q33, 2.715) (Q34, 2.711)};
\end{axis}

\node[below right,rotate=90,yshift=-9.5pt,xshift=-2.5pt,style={rred}] (Q31) {\tiny Out of Memory};
\node[below right,rotate=90,yshift=-33.5pt,xshift=-2.5pt,style={rred}] (Q32) {\tiny Out of Memory};
\node[below right,rotate=90,yshift=-57.5pt,xshift=-2.5pt,style={rred}] (Q33) {\tiny Out of Memory};
\node[below right,rotate=90,yshift=-81.5pt,xshift=-2.5pt,style={rred}] (Q34) {\tiny Out of Memory};

\node[below right,rotate=90,yshift=-14.5pt,xshift=-2.5pt,style={ggreen}] (Q31a) {\tiny Out of Memory};
\node[below right,rotate=90,yshift=-38.5pt,xshift=-2.5pt,style={ggreen}] (Q32a) {\tiny Out of Memory};
\node[below right,rotate=90,yshift=-62.5pt,xshift=-2.5pt,style={ggreen}] (Q33a) {\tiny Out of Memory};
\node[below right,rotate=90,yshift=-86.5pt,xshift=-2.5pt,style={ggreen}] (Q34a) {\tiny Out of Memory};

\node[below right,rotate=90,yshift=-19.5pt,xshift=-2.5pt,style={ppurple}] (Q31b) {\tiny Out of Memory};
\node[below right,rotate=90,yshift=-43.5pt,xshift=-2.5pt,style={ppurple}] (Q32b) {\tiny Out of Memory};
\node[below right,rotate=90,yshift=-67.5pt,xshift=-2.5pt,style={ppurple}] (Q33b) {\tiny Out of Memory};
\node[below right,rotate=90,yshift=-91.5pt,xshift=-2.5pt,style={ppurple}] (Q34b) {\tiny Out of Memory};

\end{tikzpicture}}
    \subfigure{\begin{tikzpicture}[
  every axis/.style={
    ymajorgrids = true,
    major x tick style = transparent,
    xtick = data,
    enlarge x limits=0.25,
    symbolic x coords={
      Q41,
      Q42,
      Q43,
    },
    width  = 0.28*\textwidth,
    height = 4cm,
    ylabel={Scale factor 10},
    y label style = {font=\small,at={(1.15,0.5)}},
    ybar stacked,
    ybar=1.2pt,
    ymin=0,
    ymax=9,
    scaled y ticks = false,
    bar width=4pt,
    legend cell align=left,
    legend style={
            at={(1,1.05)},
            anchor=south east,
            column sep=1ex
    },
  },
]
\begin{axis}[bar shift=-9pt]
    \addplot[style={color=bblue,fill=bblue}]
        coordinates {(Q41, 0.02) (Q42, 0.019) (Q43, 0.02)};
    \addplot[style={color=bbblue,fill=bbblue}]
        coordinates {(Q41, 7.845) (Q42, 7.368) (Q43, 3.085)};
\end{axis}

\node[below right,rotate=90,yshift=-4pt,xshift=-2.5pt,style={rred}] (Q11) at (0, 0.01) {\tiny Out of Memory};
\node[below right,rotate=90,yshift=-28pt,xshift=-2.5pt,style={rred}] (Q12) at (0, 0.01) {\tiny Out of Memory};
\node[below right,rotate=90,yshift=-52pt,xshift=-2.5pt,style={rred}] (Q13) at (0, 0.01) {\tiny Out of Memory};

\node[below right,rotate=90,yshift=-9pt,xshift=-2.5pt,style={ggreen}] (Q11a) at (0, 0.01) {\tiny Out of Memory};
\node[below right,rotate=90,yshift=-33pt,xshift=-2.5pt,style={ggreen}] (Q12b) at (0, 0.01) {\tiny Out of Memory};
\node[below right,rotate=90,yshift=-57pt,xshift=-2.5pt,style={ggreen}] (Q13c) at (0, 0.01) {\tiny Out of Memory};

\node[below right,rotate=90,yshift=-14pt,xshift=-2.5pt,style={ppurple}] (Q11aa) at (0, 0.01) {\tiny Out of Memory};
\node[below right,rotate=90,yshift=-38pt,xshift=-2.5pt,style={ppurple}] (Q12ba) at (0, 0.01) {\tiny Out of Memory};
\node[below right,rotate=90,yshift=-62pt,xshift=-2.5pt,style={ppurple}] (Q13ca) at (0, 0.01) {\tiny Out of Memory};

\end{tikzpicture}}
    \caption{Comparing the runtime performance of \frameworkname\ and the baselines}
    \label{fig:laptop_runtime_comparison}
\end{figure*}
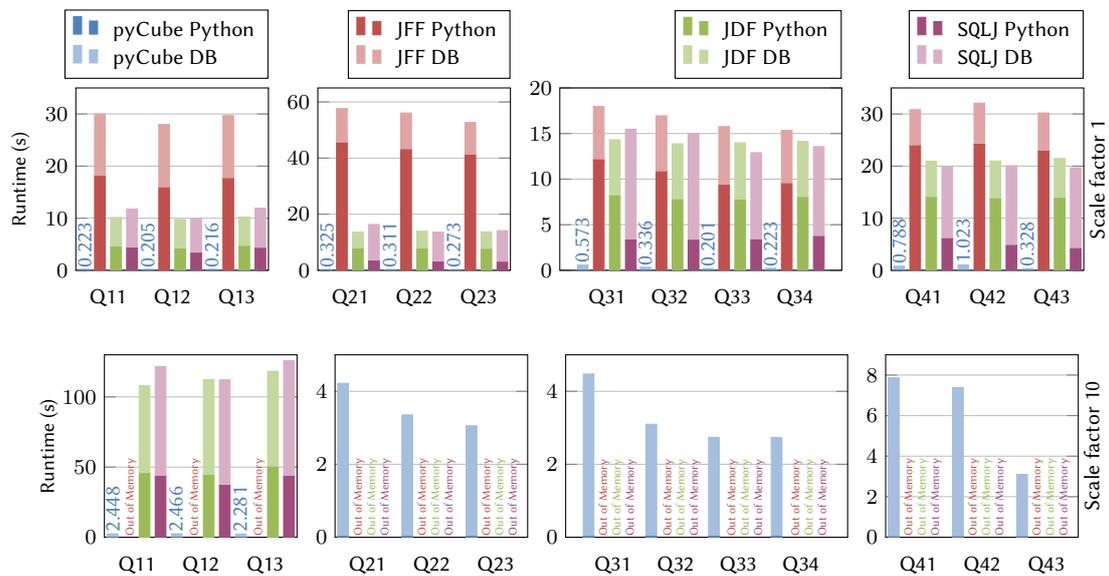

\frameworkname\ vastly outperforms all baselines for all SSB queries and is in some instances two orders of magnitude faster.
The baselines scale worse than \frameworkname\ with baseline runtimes for QF 1 ranging from 10 to 120 seconds for scale factors 1 and 10, respectively, while \frameworkname\ runtimes range from 0.2 to 2.5 seconds.
Only \frameworkname\ is able to perform all SSB queries for all scale factors without running out of memory.
JoinDimensionsFirst and \sql Join are able to perform QF 1 for scale factor 10.
Otherwise all baselines failed to perform any QF for scale factor 10.
Furthermore all baselines were unable to perform QF 4 for scale factor 5 while scale factor 2 was the highest JoinFactsFirst was able to achieve before running out of memory.
Figure~\ref{fig:laptop_runtime_comparison} shows the runtime of \frameworkname\ and the baselines for scale factors 1 and 10.
The runtime for scale factors 2 and 5 can be seen in the extended version of the paper~\cite{vang23}.
The runtime is split into time spent in Python and time spent in the database.
The time spent in the database constitutes the point in time when control has been passed on to the Python database adapter with a request until the adapter returns with a result.
The remaining time is Python time.
The runtimes for \frameworkname\ are too small to produce a noticeable bar so the combined total runtime is given where the bar should have been.
The vast majority of time in \frameworkname\ is spent in the database.
This is because \frameworkname\ only spends its Python time generating \sql\ and converting the result set to a dataframe.
The \sql\ query generated by \frameworkname\ includes all tables to be joined, all predicates in the \inlinecode{WHERE} clause and the aggregation function in the \inlinecode{SELECT} clause.
This enables the database to do some of the usual optimizations such as predicate pushdown, on-the-fly aggregation and join-order selection.

The baselines are slowed down by loading large amounts of data into memory and by filtering and reshaping the large dataframes in Python.
This can especially be seen for JoinFactsFirst where the majority of time is spent in Python as seen in Figure~\ref{fig:laptop_runtime_comparison}.
This is because the hierarchies are joined with the fact table first and as a result creates new copies of the fact data for every join.
Furthermore joins are eagerly evaluated in pandas.
JoinFactsFirst and JoinDimensionsFirst have more or less the same Python and database time split.
However, JoinDimensionsFirst is generally a lot quicker than the JoinFactsFirst since the smaller dimension tables are denormalized before joining them with the fact table.
In general \sql Join uses the same amount of time as JoinDimensionsFirst.
This is probably because only the joins in the \sql Join \sql\ query provide any information for the DBMS to optimize the query by selecting the best join ordering.
The DBMS joins all tables for \sql Join and as such more time is spent waiting for the database.
The Python and database time split is in general the same for scale factors 2, 5 and 10.

\subsection{Memory Usage}
\begin{figure*}
    \subfigure{\begin{tikzpicture}
    \begin{axis}[
        ymajorgrids = true,
        ytick={0,5,10,15,20,25},
        xtick={0,1,2,5,10},
        width = 0.26*\linewidth,
        height = 4cm,
        xlabel = {Scale Factor},
        ylabel = {Memory (GB)},
        ymin = 0, 
        ymax = 25,
        title = {QF 1},
        ]

        \addplot[pyCubeQ1Style]
            plot coordinates {(1, 0.104) (2, 0.097) (5, 0.097) (10, 0.101)};
            \addlegendentry{pycube\_q1}
        \addplot[pyCubeQ2Style]
            plot coordinates {(1, 0.102) (2, 0.099) (5, 0.097) (10, 0.101)};
            \addlegendentry{pycube\_q2}
        \addplot[pyCubeQ3Style]
            plot coordinates {(1, 0.102) (2, 0.097) (5, 0.097) (10, 0.101)};
            \addlegendentry{pycube\_q3}

        \addplot[JFFQ1Style]
            plot coordinates {(1, 10.313) (2, 20.529)};
            \addlegendentry{JFF\_q1}
        \addplot[JFFQ2Style]
            plot coordinates {(1, 8.671) (2, 17.248)};
            \addlegendentry{JFF\_q2}
        \addplot[JFFQ3Style]
            plot coordinates {(1, 10.313) (2, 20.529)};
            \addlegendentry{JFF\_q3}
        \node[xshift=4pt] (JFFOoM) at (axis cs:2, 22.5) {\tiny Out of Memory};

        \addplot[JDFQ1Style]
            plot coordinates {(1, 2.093) (2, 3.778) (5, 10.046) (10, 19.934)};
            \addlegendentry{JDF\_q1}
        \addplot[JDFQ2Style]
            plot coordinates {(1, 2.093) (2, 3.776) (5, 10.046) (10, 19.935)};
            \addlegendentry{JDF\_q2}
        \addplot[JDFQ3Style]
            plot coordinates {(1, 2.234) (2, 4.239) (5, 10.496) (10, 20.829)};
            \addlegendentry{JDF\_q3}

        \addplot[SQLJQ1Style]
            plot coordinates {(1, 2.06) (2, 3.812) (5, 10.01) (10, 19.924)};
            \addlegendentry{\sql J\_q1}
        \addplot[SQLJQ2Style]
            plot coordinates {(1, 2.093) (2, 3.806) (5, 10.044) (10, 19.934)};
            \addlegendentry{\sql J\_q2}
        \addplot[SQLJQ3Style]
            plot coordinates {(1, 2.294) (2, 4.48) (5, 11.059) (10, 22.022)};
            \addlegendentry{\sql J\_q3}
        \legend{}
    \end{axis}
\end{tikzpicture}}
    \subfigure{\begin{tikzpicture}
    \begin{axis}[
        ymajorgrids = true,
        ytick={0,5,10,15,20,25},
        xtick={0,1,2,5,10},
        width = 0.26*\linewidth,
        height = 4cm,
        xlabel = {Scale Factor},
        ymin = 0, 
        ymax = 25,
        title = {QF 2},
        ]

        \addplot[pyCubeQ1Style]
            plot coordinates {(1, 0.104) (2, 0.098) (5, 0.099) (10, 0.098)};
            \addlegendentry{pycube\_q1}
        \addplot[pyCubeQ2Style]
            plot coordinates {(1, 0.103) (2, 0.098) (5, 0.099) (10, 0.098)};
            \addlegendentry{pycube\_q2}
        \addplot[pyCubeQ3Style]
            plot coordinates {(1, 0.105) (2, 0.098) (5, 0.098) (10, 0.099)};
            \addlegendentry{pycube\_q3}

        \addplot[JFFQ1Style]
            plot coordinates {(1, 21.372)};
            \addlegendentry{JFF\_q1}
        \addplot[JFFQ2Style]
            plot coordinates {(1, 20.527)};
            \addlegendentry{JFF\_q2}
        \addplot[JFFQ3Style]
            plot coordinates {(1, 20.527)};
            \addlegendentry{JFF\_q3}
        \node[anchor = east, xshift = 3.5pt] (JFFOoM) at (axis cs:5, 23) {\tiny Out of Memory};

        \addplot[JDFQ1Style]
            plot coordinates {(1, 4.114) (2, 8.018) (5, 19.888)};
            \addlegendentry{JDF\_q1}
        \addplot[JDFQ2Style]
            plot coordinates {(1, 3.834) (2, 7.454) (5, 18.483)};
            \addlegendentry{JDF\_q2}
        \addplot[JDFQ3Style]
            plot coordinates {(1, 3.833) (2, 7.455) (5, 18.484)};
            \addlegendentry{JDF\_q3}
        \node[anchor = east, xshift = 6pt] (JDFOoM) at (axis cs:10, 17) {\tiny Out of Memory};

        \addplot[SQLJQ1Style]
            plot coordinates {(1, 3.267) (2, 6.427) (5, 16.15)};
            \addlegendentry{\sql J\_q1}
        \addplot[SQLJQ2Style]
            plot coordinates {(1, 2.794) (2, 5.477) (5, 13.551)};
            \addlegendentry{\sql J\_q2}
        \addplot[SQLJQ3Style]
            plot coordinates {(1, 2.793) (2, 5.478) (5, 13.552)};
            \addlegendentry{\sql J\_q3}
        \legend{}
    \end{axis}
\end{tikzpicture}}
    \subfigure{\begin{tikzpicture}
    \begin{axis}[
        ymajorgrids = true,
        ytick={0,5,10,15,20,25},
        xtick={0,1,2,5,10},
        width = 0.26*\linewidth,
        height = 4cm,
        xlabel = {Scale Factor},
        ymin = 0, 
        ymax = 25,
        title = {QF 3},
        ]

        \addplot[pyCubeQ1Style]
            plot coordinates {(1, 0.103) (2, 0.098) (5, 0.098) (10, 0.098)};
            \addlegendentry{pycube\_q1}
        \addplot[pyCubeQ2Style]
            plot coordinates {(1, 0.105) (2, 0.098) (5, 0.099) (10, 0.098)};
            \addlegendentry{pycube\_q2}
        \addplot[pyCubeQ3Style]
            plot coordinates {(1, 0.103) (2, 0.098) (5, 0.098) (10, 0.098)};
            \addlegendentry{pycube\_q3}
        \addplot[pyCubeQ4Style]
            plot coordinates {(1, 0.104) (2, 0.098) (5, 0.098) (10, 0.098)};
            \addlegendentry{pycube\_q4}

        \addplot[JFFQ1Style]
            plot coordinates {(1, 7.726) (2, 15.328)};
            \addlegendentry{JFF\_q1}
        \addplot[JFFQ2Style]
            plot coordinates {(1, 6.224) (2, 12.329)};
            \addlegendentry{JFF\_q2}
        \addplot[JFFQ3Style]
            plot coordinates {(1, 4.677) (2, 9.235)};
            \addlegendentry{JFF\_q3}
        \addplot[JFFQ4Style]
            plot coordinates {(1, 4.958) (2, 9.798)};
            \addlegendentry{JFF\_q4}
        \node[rotate = 50] (JFFOoM) at (axis cs: 2, 18) {\tiny Out of Memory};

        \addplot[JDFQ1Style]
            plot coordinates {(1, 4.443) (2, 8.769)};
            \addlegendentry{JDF\_q1}
        \addplot[JDFQ2Style]
            plot coordinates {(1, 4.208) (2, 8.298) (5, 20.593)};
            \addlegendentry{JDF\_q2}
        \addplot[JDFQ3Style]
            plot coordinates {(1, 3.505) (2, 6.892) (5, 17.077)};
            \addlegendentry{JDF\_q3}
        \addplot[JDFQ4Style]
            plot coordinates {(1, 3.693) (2, 7.267) (5, 18.014)};
            \addlegendentry{JDF\_q4}
        \node[anchor = east, xshift = 6pt] (JDFOoM) at (axis cs: 10, 17) {\tiny Out of Memory};

        \addplot[SQLJQ1Style]
            plot coordinates {(1, 3.98) (2, 7.851) (5, 19.487)};
            \addlegendentry{\sql J\_q1}
        \addplot[SQLJQ2Style]
            plot coordinates {(1, 3.981) (2, 7.851) (5, 19.487)};
            \addlegendentry{\sql J\_q2}
        \addplot[SQLJQ3Style]
            plot coordinates {(1, 2.792) (2, 5.477) (5, 13.551)};
            \addlegendentry{\sql J\_q3}
        \addplot[SQLJQ4Style]
            plot coordinates {(1, 3.281) (2, 6.425) (5, 16.15)};
            \addlegendentry{\sql J\_q4}
        \legend{}
    \end{axis}
\end{tikzpicture}}
    \subfigure{\begin{tikzpicture}
    \begin{axis}[
        ymajorgrids = true,
        ytick={0,5,10,15,20,25,30},
        xtick={0,1,2,5,10},
        width = 0.25*\linewidth,
        height = 4cm,
        xlabel = {Scale Factor},
        ymin = 0, 
        ymax = 30,
        title = {QF 4},
        legend cell align = left,
        legend columns = 2,
        legend style = {
            at = {(1.05, 1)},
            anchor = north west,
            font = \tiny,
            mark options = {scale = 0.5},
            row sep = -0.5pt,
        },
        ]

        \addplot[pyCubeQ1Style]
            plot coordinates {(1, 0.103) (2, 0.098) (5, 0.098) (10, 0.098)};
            \addlegendentry{pyc\_q1}
        \addplot[pyCubeQ2Style]
            plot coordinates {(1, 0.104) (2, 0.098) (5, 0.098) (10, 0.098)};
            \addlegendentry{pyc\_q2}
        \addplot[pyCubeQ3Style]
            plot coordinates {(1, 0.104) (2, 0.099) (5, 0.099) (10, 0.099)};
            \addlegendentry{pyc\_q3}
        \addlegendimage{pyCubeQ4Style}
        \addlegendentry{pyc\_q4}

        \addplot[JFFQ1Style]
            plot coordinates {(1, 12.501) (2, 24.95)};
            \addlegendentry{JFF\_q1}
        \addplot[JFFQ2Style]
            plot coordinates {(1, 11.962) (2, 23.918)};
            \addlegendentry{JFF\_q2}
        \addplot[JFFQ3Style]
            plot coordinates {(1, 12.173) (2, 24.388)};
            \addlegendentry{JFF\_q3}
        \addlegendimage{JFFQ4Style}
        \addlegendentry{JFF\_q4}
        \node[xshift = 5pt] (JFFOoM) at (axis cs: 2, 27) {\tiny Out of Memory};

        \addplot[JDFQ1Style]
            plot coordinates {(1, 6.987) (2, 13.889)};
            \addlegendentry{JDF\_q1}
        \addplot[JDFQ2Style]
            plot coordinates {(1, 7.177) (2, 14.263)};
            \addlegendentry{JDF\_q2}
        \addplot[JDFQ3Style]
            plot coordinates {(1, 7.201) (2, 14.358)};
            \addlegendentry{JDF\_q3}
        \addlegendimage{JDFQ4Style}
        \addlegendentry{JDF\_q4}
        \node[xshift = 1pt, yshift = -2pt] (JDFOoM) at (axis cs: 5, 12.5) {\tiny Out of Memory};

        \addplot[SQLJQ1Style]
            plot coordinates {(1, 4.543) (2, 8.974)};
            \addlegendentry{\sql J\_q1}
        \addplot[SQLJQ2Style]
            plot coordinates {(1, 4.907) (2, 9.705)};
            \addlegendentry{\sql J\_q2}
        \addplot[SQLJQ3Style]
            plot coordinates {(1, 4.908) (2, 9.706)};
            \addlegendentry{\sql J\_q3}
        \addlegendimage{SQLJQ4Style}
        \addlegendentry{\sql J\_q4}

    \end{axis}
\end{tikzpicture}}
    \caption{Laptop memory growth over scale factors}
    \label{fig:laptop_memory_comparison}
\end{figure*}
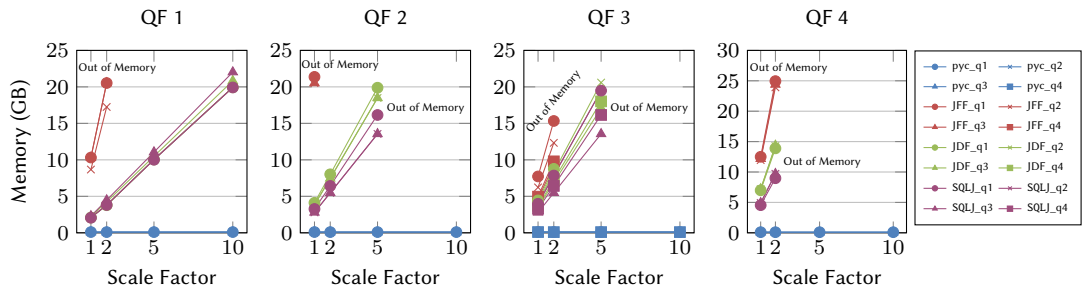

Figure~\ref{fig:laptop_memory_comparison} shows the memory usage of \frameworkname\ and the baselines measured using the \inlinecode{time} tool.
The memory shown is the resident set size of a process in GB which includes all stack and heap memory of a process in addition to the shared libraries in memory.
It does not include swapped out memory.
The charts show the same pattern found in Figure~\ref{fig:laptop_runtime_comparison} with \frameworkname\ outperforming the baselines across all queries for all scale factors.
The baselines' memory scales linearly with the scale factor while \frameworkname 's memory remains constant.
As a result JoinFactsFirst runs out of memory on scale factors 5, 2 and 5 for QFs 1, 2 and 3, respectively.
JoinDimensionsFirst and \sql Join run out of memory on scale factor 10 for QFs 2 and 3 while all baselines run out of memory on scale factor 5 for QF 4.
The experiments were repeated on a high-end server running Ubuntu, with 16 cores of AMD Epyc 7302P processor running at a 3.2GHz clock frequency and 264GB RAM to see how much memory the baselines needed in order to perform the SSB queries on higher scale factors.
JoinFactsFirst uses the most memory with the highest being about 200GB on QF 2 and in general approaches 100GB on scale factor 10 for all QFs.
JoinDimensionsFirst and \sql Join's memory ranges between 30GB and 70GB for scale factor 10 on QFs 2 to 4.
These queries cannot fit on laptops that data scientists use.
The figure can be seen in the extended version of the paper~\cite{vang23}.

\subsection{Code Comparison}
Listing~\ref{lst:query41pyCube} shows SSB query 4.1 implemented using \frameworkname.
SSB query 4.1 implemented with the JDF baseline can be seen in the extended version of the paper~\cite{vang23}.
\begin{lstlisting}[caption={Query 4.1 of SSB in pyCube},label={lst:query41pyCube}]
view.columns(view.date.year.y_year.members()) \
    .rows(view.customer.nation.n_nation.members()) \
    .where(
    (view.customer.region.r_region == "AMERICA")
    & (view.supplier.region.r_region == "AMERICA")
    & (
            (view.part.mfgr.m_mfgr == "MFGR#1")
            | (view.part.mfgr.m_mfgr == "MFGR#2")
      )
    ) \
    .measures(profit
              =view.lo_revenue - view.lo_supplycost)\
    .output()
\end{lstlisting}
Table~\ref{tab:code_comparison} compares the sizes of the implementations.
\begin{table}
    \caption{Size differences between \frameworkname\ and the JDF baseline}
    \begin{tabular}{@{}cccc@{}}
        \toprule
        & \frameworkname & JDF & JDF reduced \\ 
        \midrule
        Characters & 412 & 2725 & 2571 \\
        Statements & 1 & 32 & 22 \\
        Tables & 13 & 13 & 13 \\
        \bottomrule
    \end{tabular}
    \label{tab:code_comparison}
\end{table}
The definition of a statement in the grammar specification for Python~\cite{pythonGrammar} is used.
The \frameworkname\ query has 6.6 times fewer characters than the JDF query and 31 fewer statements.
The statement comparison may seem skewed since expressions in \frameworkname\ are chained into a single statement.
However \frameworkname\ allows expressions to be chained while maintaining readability, which pandas only can do to a limited extent.
JDF reduced in Table~\ref{tab:code_comparison} is the smallest form of the JDF query without making the implementation overly convoluted.
The reduction is achieved by chaining certain \inlinecode{merge} methods.
Even with the reduction, the \frameworkname\ query still has 6.2 times fewer characters and 21 fewer statements.
The reduced JDF query can be seen in the extended version of the paper~\cite{vang23}.
A major reason for the increased complexity of reading and writing the JDF query is due to the higher user responsibilities in pandas than in \frameworkname.
In pandas, the user needs to manually fetch the data from the database and then merge, filter and pivot the dataframes correctly.
In fact, 13 of the 32 statements in the JDF query come from the many tables that need to be fetched from the database.
This is because the data is structured in a snowflake schema.
However, even if the data was structured in a star schema, there still would need to be five \lstinline[breaklines=false]{read_sql} statements (four dimensions and one fact table) and four invocations of the \inlinecode{merge} method just for loading the data correctly into memory.
The \inlinecode{read_sql} and \inlinecode{merge} methods are well designed and easy to write but writing nine invocations of two methods back-to-back is repetitive which will increase the chances of error from the user.
Additionally, when merging dataframes, all relevant keys must be provided and column names need to be given suffixes in case of column name clashes.
In contrast, the user responsibility in \frameworkname\ is specifying the metadata and values of the desired dataframe using methods with meaningful names and then calling \inlinecode{output} as is shown in Listing~\ref{lst:query41pyCube}.
This results in significantly more compact code that is easier to read, write, debug and maintain.

\section{Conclusion and Future Work}\label{sec:conclusion}
This paper has presented \frameworkname: a Python-based data cube tool.
\frameworkname\ has been designed to match the suite of tools usually employed by data scientists.
The user interface of \frameworkname\ has been shown alongside how \frameworkname\ processes user queries.
\frameworkname\ has been experimentally evaluated.
The results show that \frameworkname\ outperforms pandas both in runtime and in memory for data cube analysis.
Future work includes expanding \frameworkname: (1) to be able to manage and query data cubes using data in distributed flat files, (2) by providing more ways to handle metadata and (3) by handling schema updates in a graceful manner.

\bibliography{/home/sigmundur/work/aau/references/ref}

\appendix

\section{Cube Inference Algorithm}\label{sec:inf_algo}
The algorithm for inferring the cube assumes that the underlying relational data source is structured in a snowflake schema and uses this structure in four steps: (1) discover the fact table, (2) discover the lowest level of all dimensions, (3) construct the hierarchies, and (4) discover the measures.
The metadata of the cube is described with an RDF graph using the QB4OLAP~\cite{etcheverry12} vocabulary.
The algorithm is shown in Algorithm~\ref{alg:inf_cube}.
\begin{algorithm}
    \caption{Inferring the Cube}
    \label{alg:inf_cube}
    \begin{algorithmic}[1]
        \State factTable = \Call{FindFactTable}{ }\label{algl:findFactTable}
        \State bottomLevels = \Call{FindBottomLevels}{factTable}\label{algl:findBottomLevels}
        \State allLevels = []\label{algl:allLevels}
        \State levelAttributes = []\label{algl:levelAttributes}
        \ForAll{level $\in$ bottomLevels}\label{algl:forloop}
            \State levels = []\label{algl:levels}
            \State level\_attributes = []\label{algl:level_attributes}
            \State currentLevel = level\label{algl:currentLevel}
            \While{currentLevel is not Null}\label{algl:whileloop}
                \State levels.append(currentLevel)\label{algl:append}
                \State c = \Call{FindNonKeyColumns}{currentLevel}\label{algl:findNonKeyColumns}
                \If{len(c) > 1}\label{algl:ifLen}
                    \State attributes = \Call{FindLevelAttributes}{c}\label{algl:removeLevelMember}
                    \State level\_attributes.append(attributes)
                \EndIf
                \State currentLevel=\Call{FindNextTable}{currentLevel}\label{algl:findNextTable}
            \EndWhile
            \State allLevels.append(levels)
            \State levelAttributes.append(level\_attributes)
        \EndFor
        \State measures = \Call{FindMeasures}{factTable}\label{algl:findMeasures}
        \State \Call{AddToGraph}{levels, levelAttributes, measures}\label{algl:addToGraph}
    \end{algorithmic}
\end{algorithm}

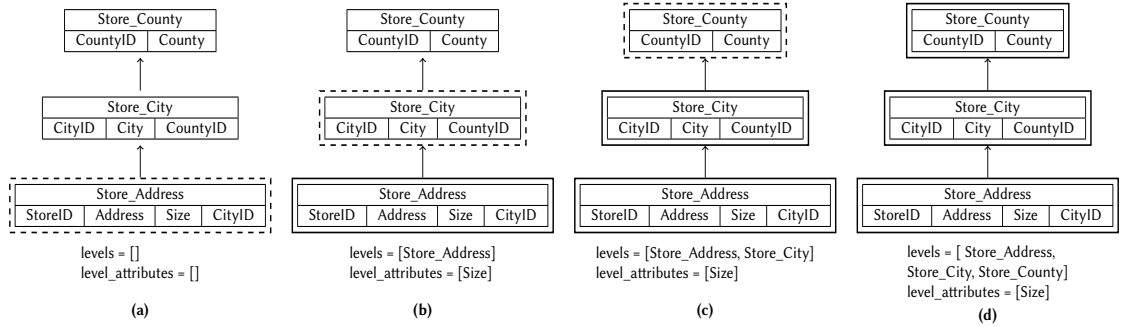
\begin{figure*}
    \resizebox{\linewidth}{15em}{
        \begin{tikzpicture}
            \usetikzlibrary{shapes}
            \tikzstyle{every node}=[]
            \node[align=left] (stage1) at (0, -1) {
                levels = [] \\
                level\_attributes = []
            };
            \node (n4) at (0, -1.8) {
                \textbf{(a)}
            };
            \node[thick, draw=black, rectangle, dashed] (store1) at (0,0) {
              \begin{tabular}{|c|c|c|c|}
              \hline
              \multicolumn{4}{|c|}{Store\_Address} \\
              \hline
              StoreID & Address & Size & CityID \\
              \hline
              \end{tabular}
            };
        
            \node (store_city1) at (0,1.5) {
              \begin{tabular}{|c|c|c|}
              \hline
              \multicolumn{3}{|c|}{Store\_City} \\
              \hline
              CityID & City & CountyID \\
              \hline
              \end{tabular}
            };
        
            \node (store_county1) at (0,3) {
              \begin{tabular}{|c|c|}
              \hline
              \multicolumn{2}{|c|}{Store\_County} \\
              \hline
              CountyID & County \\
              \hline
              \end{tabular}
            };
        
            \draw[->] (store1.north) to (store_city1.south);
            \draw[->] (store_city1.north) to (store_county1.south);
        
            \node[align=left] (stage2) at (5.5, -1) {
                levels = [Store\_Address] \\
                level\_attributes = [Size]
            };
            \node (n4) at (5.5, -1.8) {
                \textbf{(b)}
            };
            \node[thick, draw=black, rectangle] (store2) at (5.5,0) {
              \begin{tabular}{|c|c|c|c|}
              \hline
              \multicolumn{4}{|c|}{Store\_Address} \\
              \hline
              StoreID & Address & Size & CityID \\
              \hline
              \end{tabular}
            };
        
            \node[thick, draw=black, rectangle, dashed] (store_city2) at (5.5,1.5) {
              \begin{tabular}{|c|c|c|}
              \hline
              \multicolumn{3}{|c|}{Store\_City} \\
              \hline
              CityID & City & CountyID \\
              \hline
              \end{tabular}
            };
        
            \node[] (store_county2) at (5.5,3) {
              \begin{tabular}{|c|c|}
              \hline
              \multicolumn{2}{|c|}{Store\_County} \\
              \hline
              CountyID & County \\
              \hline
              \end{tabular}
            };
        
            \draw[->] (store2.north) to (store_city2.south);
            \draw[->] (store_city2.north) to (store_county2.south);
        
            \node[align=left] (stage3) at (11.0, -1) {
                levels = [Store\_Address, Store\_City] \\
                level\_attributes = [Size]
            };
            \node (n4) at (11.0, -1.8) {
                \textbf{(c)}
            };
            \node[thick, draw=black, rectangle] (store2) at (11.0,0) {
              \begin{tabular}{|c|c|c|c|}
              \hline
              \multicolumn{4}{|c|}{Store\_Address} \\
              \hline
              StoreID & Address & Size & CityID \\
              \hline
              \end{tabular}
            };
        
            \node[thick, draw=black, rectangle] (store_city2) at (11.0,1.5) {
              \begin{tabular}{|c|c|c|}
              \hline
              \multicolumn{3}{|c|}{Store\_City} \\
              \hline
              CityID & City & CountyID \\
              \hline
              \end{tabular}
            };
        
            \node[thick, draw=black, rectangle, dashed] (store_county2) at (11.0,3) {
              \begin{tabular}{|c|c|}
              \hline
              \multicolumn{2}{|c|}{Store\_County} \\
              \hline
              CountyID & County \\
              \hline
              \end{tabular}
            };
        
            \draw[->] (store2.north) to (store_city2.south);
            \draw[->] (store_city2.north) to (store_county2.south);
        
            \node (n4) at (16.5, -1.9) {
                \textbf{(d)}
            };
        
            \node[align=left] (stage4) at (16.5, -1.15) {
                levels = [ Store\_Address, \\
                Store\_City, Store\_County] \\
                level\_attributes = [Size]
            };
        
            \node[thick, draw=black, rectangle] (store2) at (16.5,0) {
              \begin{tabular}{|c|c|c|c|}
              \hline
              \multicolumn{4}{|c|}{Store\_Address} \\
              \hline
              StoreID & Address & Size & CityID \\
              \hline
              \end{tabular}
            };
        
            \node[thick, draw=black, rectangle] (store_city2) at (16.5,1.5) {
              \begin{tabular}{|c|c|c|}
              \hline
              \multicolumn{3}{|c|}{Store\_City} \\
              \hline
              CityID & City & CountyID \\
              \hline
              \end{tabular}
            };
        
            \node[thick, draw=black, rectangle] (store_county2) at (16.5,3) {
              \begin{tabular}{|c|c|}
              \hline
              \multicolumn{2}{|c|}{Store\_County} \\
              \hline
              CountyID & County \\
              \hline
              \end{tabular}
            };
        
            \draw[->] (store2.north) to (store_city2.south);
            \draw[->] (store_city2.north) to (store_county2.south);
        \end{tikzpicture}
    }
    \caption{Constructing the hierarchy in the Store dimension}
    \label{fig:hierarchy_example}
\end{figure*}

The \inlinecode{FindFactTable} procedure discovers the fact table on Line~\ref{algl:findFactTable} by using the system catalogs of the data source to find the table with the largest cardinality, since fact tables usually are much larger than the other tables. 
The \inlinecode{FindBottomLevels} procedure finds all bottoms levels on Line~\ref{algl:findBottomLevels} in every hierarchy by following the FKs contained in the fact table.
The \inlinecode{allLevels} and \inlinecode{levelAttributes} variables on Line~\ref{algl:allLevels}~and~\ref{algl:levelAttributes} are arrays consisting of arrays that will contain the levels and level attributes respectively for all hierarchies.
After initializing the variables, Algorithm~\ref{alg:inf_cube} constructs the hierarchies by looping through the levels in \texttt{bottomLevels} in the loop at Line~\ref{algl:forloop}.
The \texttt{levels} and \texttt{level\_attributes} variables on Line~\ref{algl:levels}~and~\ref{algl:level_attributes} will contain the levels and level attributes respectively for each hierarchy.
On Line~\ref{algl:whileloop} the algorithm constructs an individual hierarchy by traversing the snowflake schema from the lowest level in the hierarchy until the \inlinecode{FindNextTable} procedure cannot find another table.
An example which shows the construction of the hierarchy in the \texttt{store} dimension is shown in Figure~\ref{fig:hierarchy_example}.

The example shows the state of the \inlinecode{levels} and \inlinecode{level\_attributes} in each iteration of the while loop, while the previous and current values of \inlinecode{currentLevel} are indicated by the surrounding solid and dashed border respectively.
In Figure~\ref{fig:hierarchy_example}a the \inlinecode{levels} and \inlinecode{level\_attributes} arrays are empty, while \inlinecode{currentLevel} is set to \inlinecode{Store\_Address}.
\inlinecode{Store\_Address} is appended to the \inlinecode{levels} array on Line~\ref{algl:append} since \inlinecode{Store\_Address} is not \inlinecode{Null}.
Then every non-key column of \inlinecode{Store\_Address} is found using the \inlinecode{FindNonKeyColumns} procedure on Line~\ref{algl:findNonKeyColumns} and stored in the variable \inlinecode{c}.
Since the \inlinecode{StoreID} and \inlinecode{CityID} columns are the PK and FK, respectively, of the table, \inlinecode{FindNonKeyColumns} returns the \inlinecode{Address} and \inlinecode{Size} columns.
Generally if there are non-key columns in a level table, the level member will always be considered to be one of them.
Level attributes are therefore only contained in the \inlinecode{c} variable if its length is greater than one, in which case \inlinecode{c} contains one level member and one or more level attributes.
The level attributes are retrieved by the \inlinecode{FindLevelAttribute} procedure, which removes the level member and returns the remaining elements in \inlinecode{c}.
Removing the level member is done by removing the column whose name is most similar to the table name.
Many string comparison algorithms exists~\cite{navarro01}.
We are using the Levenshtein distance. 
In this case \inlinecode{Address} is removed, since it is more similar to \inlinecode{Store\_Address} than \inlinecode{Size} is.

The \inlinecode{FindNextTable} procedure on Line~\ref{algl:findNextTable} finds the next table by following the FK, if there exists one, contained in \inlinecode{currentLevel}.
If no FK is in \inlinecode{currentLevel}, then \inlinecode{FindNextTable} returns \inlinecode{Null}.
Since \inlinecode{Store\_Address} contains a FK \lstinline[breaklines=false]{FindNextTable} returns \inlinecode{Store\_City}, which is assigned to the \inlinecode{currentLevel} variable and a new iteration of the while loop is started.
Figure~\ref{fig:hierarchy_example}b shows the state of the algorithm at the beginning of the second iteration, while Figure~\ref{fig:hierarchy_example}c~and~\ref{fig:hierarchy_example}d show the state of the third and fourth iteration respectively.
When all levels and level attributes have been found for a particular dimension, they are appended to the \inlinecode{allLevels} and \inlinecode{levelAttributes} arrays respectively.

The measures are discovered using the \inlinecode{FindMeasures} procedure on Line~\ref{algl:findMeasures}, when all levels and level attributes have been found for all dimensions.
Since the fact table is assumed to only contain numerical values and FKs, which are surrogate keys that reference the lowest level in a dimension, the measures are found by finding all non-key numerical columns in the fact table.
The \inlinecode{SUM} aggregate function is applied to all measures.
Finally the metadata is added to the RDF graph on Line~\ref{algl:addToGraph} using the \inlinecode{AddToGraph} procedure.
Listing~\ref{lst:store_dimension} shows the metadata of the \inlinecode{Store} dimension, while Listing~\ref{lst:dsd} shows the components that are added to the \inlinecode{DataStructureDefinition}, when running Algorithm~\ref{alg:inf_cube} on the entire running example.
Both example use QB4OLAP serialized in the turtle format and prefixes are omitted for the sake of space.

\begin{lstlisting}[label={lst:store_dimension},caption={The store dimension in QB4OLAP}]
eg:store a qb:DimensionProperty .

eg:size a qb:AttributeProperty .

eg:store_address a qb4o:LevelProperty ;
    qb4o:hasAttribute eg:size ;
    qb4o:inDimension eg:store ;
    qb4o:parentLevel eg:store_city .

eg:store_city a qb4o:LevelProperty ;
    qb4o:inDimension eg:store ;
    qb4o:parentLevel eg:store_county .

eg:store_county a qb4o:LevelProperty ;
    qb4o:inDimension eg:store .

\end{lstlisting}

\begin{lstlisting}[label={lst:dsd},caption={The DataStructureDefinition produced by running Algorithm~\protect\ref{alg:inf_cube} on the running example}]
eg:salesdb_snowflake_dsd a qb:DataStructureDefinition ;
    qb:component [ qb4o:level eg:supplier_name ],
        [ qb4o:level eg:store_address ],
        [ qb4o:level eg:product_name ],
        [ qb4o:level eg:date_day ],
        [ qb:measure eg:unit_sales ;
            qb4o:hasAggregateFunction qb4o:sum ],
        [ qb:measure eg:total_sales_price ;
            qb4o:hasAggregateFunction qb4o:sum ] .
\end{lstlisting}

\section{Generating the SQL query}
The \inlinecode{get_from_clause_subset(}$d$\inlinecode{)} is a function which returns, when given a dimension $d$, the query subset needed for the \texttt{FROM} clause for $d$.
An example result is given in Listing~\ref{lst:get_from_clause_subset_example}.
The notation $\ell$\inlinecode{.pk} and $\ell$\inlinecode{.fk} indicates the primary and foreign key of level schema $\ell$.
The fact table is denoted as \inlinecode{ft} and \inlinecode{ft.}$\ell$ denotes the foreign key on the fact table referencing $\ell$.
The lowest level in dimension $d$ is indicated by $\ell_d$ while $\ell_i^j$ denotes the level $j$ steps above the $\ell_i$ level in the dimension hierarchy, i.e., $\ell_i\preceq^j\ell_i^j$.
The number of steps needed to go from $\ell_d$ to the highest level in the dimension hierarchy is $h$.

\begin{lstlisting}[caption={Example \inlinecode{get_from_clause_subset(}$d$\inlinecode{)} result},label={lst:get_from_clause_subset_example},mathescape=true,lineskip=4pt]
JOIN $\ell_d$ ON $\ell_d$.pk = ft.$\ell_d$
JOIN $\ell_d^1$ ON $\ell_d^1$.pk = $\ell_d$.fk
JOIN $\ell_d^2$ ON $\ell_d^2$.pk = $\ell_d^1$.fk
$\dots$
JOIN $\ell^{h-1}$ ON $\ell^{h-1}$.pk = $\ell^{h-2}$.fk
JOIN $\ell^{h}$ ON $\ell^{h}$.pk = $\ell^{h-1}$.fk
\end{lstlisting}

Let $\ell_{p_i}$, $a_{p_i}$ and $lit_{p_i}$ be the level, attribute and literal of predicate $p_i$ in $p$.
The \inlinecode{inclusion_where_clause_subset} method places all attributes used in the axes in a series of inclusion conditions while the \inlinecode{predicate_where_clause_subset(}$p$\inlinecode{)} method places all predicates specified in $p$.
The \inlinecode{inclusion_where_clause_subset(}$AX$\inlinecode{)} and \inlinecode{predicate_where_clause_subset(}$p$\inlinecode{)} methods are shown in Listings~\ref{lst:inclusion_where_clause_subset_example} and \ref{lst:predicate_where_clause_subset_example}.

\begin{lstlisting}[caption={The \inlinecode{inclusion_where_clause_subset(}$AX$\inlinecode{)} method},label={lst:inclusion_where_clause_subset_example},mathescape=true,lineskip=4pt]
$\ell_1$.$a_1$ IN ($lm_1$) AND 
$\ell_2$.$a_2$ IN ($lm_2$) AND 
$\cdots$ 
$\ell_n$.$a_n$ IN ($lm_n$)
\end{lstlisting}

\begin{lstlisting}[caption={The \inlinecode{predicate_where_clause_subset(}$p$\inlinecode{)} method},label={lst:predicate_where_clause_subset_example},mathescape=true,lineskip=4pt]
$\ell_{p_1}$.$a_{p_1}$ $\omega_1$ $lit_{p_1}$ $\otimes$ 
$\ell_{p_2}$.$a_{p_2}$ $\omega_2$ $lit_{p_2}$ $\otimes$
$\dots$
$\ell_{p_o}$.$a_{p_o}$ $\omega_o$ $lit_{p_o}$
\end{lstlisting}
If two predicates are joined with a logical and then $\otimes$ is \texttt{AND}.
Conversely $\otimes$ is \texttt{OR} if two predicates are joined with a logical or.
Parentheses are placed according to the evaluation order.

An full example of the generated \sql, after invoking the \inlinecode{output} method on the view in Section~\ref{sec:sql_generation}, is shown in Listing~\ref{lst:sql_query_example}.
\begin{lstlisting}[caption={Resulting \sql\ statement},label={lst:sql_query_example}]
SELECT 
    month.month, 
    category.category, 
    city.city, 
    SUM(ft.TotalSalesPrice) AS TotalSalesPrice, 
    SUM(ft.UnitSales) AS UnitSales
FROM
    ft 
    -- Date dimension join
    JOIN date ON date.datekey = fact_table.datekey
    JOIN month ON month.monthkey = date.monthkey
    JOIN year ON year.yearkey = month.yearkey

    -- Product dimension join
    JOIN part ON part.partkey = fact_table.partkey
    JOIN brand1 ON brand1.brand1key = part.brand1key
    JOIN category ON category.categorykey 
                     = brand1.categorykey
    JOIN mfgr ON mfgr.mfgrkey = category.mfgrkey

    -- Store dimension join
    JOIN city ON city.citykey = fact_table.citykey
    JOIN nation ON nation.nationkey = city.nationkey
    JOIN region ON region.regionkey = nation.regionkey
WHERE 
    -- Result of inclusion_where_clause_subset($AX$)
        month.month IN (<all months in a year>) 
    AND category.category IN ("Blouse", "Pants") 
    AND city.city IN ("Aalborg") 

    -- Result of predicate_where_clause_subset($p$)
    AND (month.month = "January" OR 
         month.month = "February")
GROUP BY
    month.month, month.monthkey,
    category.category, category.categorykey,
    city.city, city.citykey
\end{lstlisting}
\todo[inline]{Improve example by including a dimension in the predicate which is not used in the representational methods}
\todo[inline]{chr: hvad betyder det? Tager det højde for parenteser, som brugeren har angivet? (til Parentheses are placed according to the evaluation order)}

\section{Evaluation}
\subsection{Data Retrival Speeds}
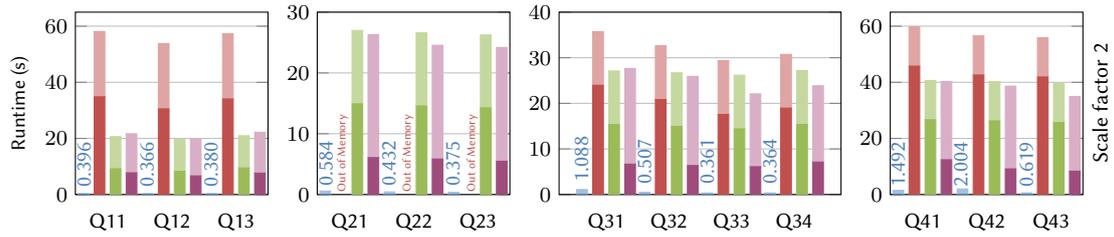
\begin{figure*}
    \subfigure{\begin{tikzpicture}[
  every axis/.style={
    ymajorgrids = true,
    major x tick style = transparent,
    xtick = data,
    enlarge x limits=0.25,
    symbolic x coords={Q11,Q12,Q13},
    width  = 0.28*\textwidth,
    height = 4cm,
    ylabel = {Runtime (s)},
    y label style = {font=\footnotesize,at={(-0.2,0.5)}},
    ybar stacked,
    ybar=1.2pt,
    ymin=0,
    ymax=65,
    scaled y ticks = false,
    bar width=4pt,
    legend cell align=left,
    legend style={
            at={(1,1.05)},
            anchor=south east,
            column sep=1ex
    },
  },
]

\begin{axis}[bar shift=-9pt]
    \addplot[style={color=bblue,fill=bblue}]
        coordinates {(Q11, 0.008) (Q12, 0.008) (Q13, 0.008)};
    \addplot[style={color=bbblue,fill=bbblue}]
        coordinates {(Q11, 0.388) (Q12, 0.358) (Q13, 0.372)};
\end{axis}

\node[below right,rotate=90,yshift=3pt,xshift=0pt,style={bblue}] (Q11) at (0.03, 0) {0.396};
\node[below right,rotate=90,yshift=-20.5pt,xshift=0pt,style={bblue}] (Q12) at (0.03, 0) {0.366};
\node[below right,rotate=90,yshift=-44.5pt,xshift=0pt,style={bblue}] (Q13) at (0.03, 0) {0.380};

\begin{axis}[bar shift=-3pt,hide axis]
    \addplot+[style={color=rred,fill=rred}]
        coordinates {(Q11, 34.827) (Q12, 30.527) (Q13, 34.069)};
    \addplot+[style={color=rrred,fill=rrred}]
        coordinates {(Q11, 23.308) (Q12, 23.271) (Q13, 23.232)};
\end{axis}

\begin{axis}[bar shift=3pt,hide axis]
    \addplot+[style={color=ggreen,fill=ggreen}]
        coordinates {(Q11, 9.17) (Q12, 8.238) (Q13, 9.483)};
    \addplot+[style={color=gggreen,fill=gggreen}]
        coordinates {(Q11, 11.471) (Q12, 11.484) (Q13, 11.474)};
\end{axis}

\begin{axis}[bar shift=9pt,hide axis]
    \addplot+[style={color=ppurple,fill=ppurple}]
        coordinates {(Q11, 7.748) (Q12, 6.652) (Q13, 7.637)};
    \addplot+[style={color=pppurple,fill=pppurple}]
        coordinates {(Q11, 13.917) (Q12, 13.116) (Q13, 14.586)};
\end{axis}
\end{tikzpicture}}
    \subfigure{\begin{tikzpicture}[
  every axis/.style={
    ymajorgrids = true,
    major x tick style = transparent,
    xtick = data,
    enlarge x limits=0.25,
    symbolic x coords={
      Q21,
      Q22,
      Q23,
    },
    width  = 0.28*\textwidth,
    height = 4cm,
    y label style = {font=\footnotesize,at={(-0.05,0.5)}},
    ybar stacked,
    ybar=1.2pt,
    ymin=0,
    ymax=30,
    scaled y ticks = false,
    bar width=4pt,
    legend cell align=left,
    legend style={
            at={(1,1.05)},
            anchor=south east,
            column sep=1ex
    },
  },
]
\begin{axis}[bar shift=-9pt]
    \addplot[style={color=bblue,fill=bblue}]
        coordinates {(Q21, 0.016) (Q22, 0.015) (Q23, 0.015)};
    \addplot[style={color=bbblue,fill=bbblue}]
        coordinates {(Q21, 0.568) (Q22, 0.417) (Q23, 0.36)};
\end{axis}

\node[below right,rotate=90,yshift=3pt,xshift=0pt,style={bblue}] (Q21) at (0.03, 0.01) {0.584};
\node[below right,rotate=90,yshift=-20.5pt,xshift=0pt,style={bblue}] (Q22) at (0.03, 0.01) {0.432};
\node[below right,rotate=90,yshift=-44.5pt,xshift=0pt,style={bblue}] (Q23) at (0.03, 0.01) {0.375};

\begin{axis}[bar shift=3pt,hide axis]
    \addplot+[style={color=ggreen,fill=ggreen}]
        coordinates {(Q21, 14.916) (Q22, 14.572) (Q23, 14.249)};
    \addplot+[style={color=gggreen,fill=gggreen}]
        coordinates {(Q21, 12.075) (Q22, 12.027) (Q23, 12.004)};
\end{axis}

\node[below right,rotate=90,yshift=-4.5pt,xshift=-2pt,style={rred}] (Q21OoM) at (0, 0.01) {\tiny Out of Memory};
\node[below right,rotate=90,yshift=-29pt,xshift=-2pt,style={rred}] (Q22OoM) at (0, 0.01) {\tiny Out of Memory};
\node[below right,rotate=90,yshift=-53pt,xshift=-2pt,style={rred}] (Q22OoM) at (0, 0.01) {\tiny Out of Memory};

\begin{axis}[bar shift=9pt,hide axis]
    \addplot+[style={color=ppurple,fill=ppurple}]
        coordinates {(Q21, 6.072) (Q22, 5.839) (Q23, 5.48)};
    \addplot+[style={color=pppurple,fill=pppurple}]
        coordinates {(Q21, 20.231) (Q22, 18.72) (Q23, 18.719)};
\end{axis}
\end{tikzpicture}}
    \subfigure{\begin{tikzpicture}[
  every axis/.style={
    ymajorgrids = true,
    major x tick style = transparent,
    xtick = data,
    enlarge x limits=0.25,
    symbolic x coords={
      Q31,
      Q32,
      Q33,
      Q34,
    },
    width  = 0.36*\textwidth,
    height = 4cm,
    y label style = {font=\footnotesize,at={(-0.05,0.5)}},
    ybar stacked,
    ybar=1.2pt,
    ymin=0,
    ymax=40,
    scaled y ticks = false,
    bar width=4pt,
    legend cell align=left,
    legend style={
            at={(1,1.05)},
            anchor=south east,
            column sep=1ex
    },
  },
]
\begin{axis}[bar shift=-9pt]
    \addplot[style={color=bblue,fill=bblue}]
        coordinates {(Q31, 0.017) (Q32, 0.018) (Q33, 0.018) (Q34, 0.019)};
    \addplot[style={color=bbblue,fill=bbblue}]
        coordinates {(Q31, 1.071) (Q32, 0.489) (Q33, 0.343) (Q34, 0.345)};
\end{axis}

\node[below right,rotate=90,yshift=-3pt,xshift=2pt,style={bblue}] (Q31) {1.088};
\node[below right,rotate=90,yshift=-26.5pt,xshift=1pt,style={bblue}] (Q32) {0.507};
\node[below right,rotate=90,yshift=-50pt,xshift=1pt,style={bblue}] (Q33) {0.361};
\node[below right,rotate=90,yshift=-73.5pt,xshift=1pt,style={bblue}] (Q34) {0.364};

\begin{axis}[bar shift=-3pt,hide axis]
    \addplot+[style={color=rred,fill=rred}]
        coordinates {(Q31, 23.967) (Q32, 20.841) (Q33, 17.563) (Q34, 18.952)};
    \addplot+[style={color=rrred,fill=rrred}]
        coordinates {(Q31, 11.775) (Q32, 11.788) (Q33, 11.791) (Q34, 11.784)};
\end{axis}
\begin{axis}[bar shift=3pt,hide axis]
    \addplot+[style={color=ggreen,fill=ggreen}]
        coordinates {(Q31, 15.334) (Q32, 14.929) (Q33, 14.398) (Q34, 15.38)};
    \addplot+[style={color=gggreen,fill=gggreen}]
        coordinates {(Q31, 11.77) (Q32, 11.767) (Q33, 11.777) (Q34, 11.806)};
\end{axis}
\begin{axis}[bar shift=9pt,hide axis]
    \addplot+[style={color=ppurple,fill=ppurple}]
        coordinates {(Q31, 6.673) (Q32, 6.358) (Q33, 6.097) (Q34, 7.132)};
    \addplot+[style={color=pppurple,fill=pppurple}]
        coordinates {(Q31, 20.971) (Q32, 19.561) (Q33, 15.983) (Q34, 16.735)};
\end{axis}
\end{tikzpicture}}
    \subfigure{\begin{tikzpicture}[
  every axis/.style={
    ymajorgrids = true,
    major x tick style = transparent,
    xtick = data,
    enlarge x limits=0.25,
    symbolic x coords={
      Q41,
      Q42,
      Q43,
    },
    width  = 0.28*\textwidth,
    height = 4cm,
    ylabel={Scale factor 2},
    y label style = {font=\small,at={(1.17,0.5)}},
    ybar stacked,
    ybar=1.2pt,
    ymin=0,
    ymax=65,
    scaled y ticks = false,
    bar width=4pt,
    legend cell align=left,
    legend style={
            at={(1,1.05)},
            anchor=south east,
            column sep=1ex
    },
  },
]
\begin{axis}[bar shift=-9pt]
    \addplot[style={color=bblue,fill=bblue}]
        coordinates {(Q41, 0.015) (Q42, 0.018) (Q43, 0.018)};
    \addplot[style={color=bbblue,fill=bbblue}]
        coordinates {(Q41, 1.477) (Q42, 1.986) (Q43, 0.601)};
\end{axis}

\node[below right,rotate=90,yshift=2.5pt,xshift=0pt,style={bblue}] (Q41) {1.492};
\node[below right,rotate=90,yshift=-21pt,xshift=0pt,style={bblue}] (Q42) {2.004};
\node[below right,rotate=90,yshift=-45.5pt,xshift=0pt,style={bblue}] (Q43) {0.619};

\begin{axis}[bar shift=-3pt,hide axis]
    \addplot+[style={color=rred,fill=rred}]
        coordinates {(Q41, 45.765) (Q42, 42.599) (Q43, 41.898)};
    \addplot+[style={color=rrred,fill=rrred}]
        coordinates {(Q41, 13.996) (Q42, 13.967) (Q43, 13.971)};
\end{axis}
\begin{axis}[bar shift=3pt,hide axis]
    \addplot+[style={color=ggreen,fill=ggreen}]
        coordinates {(Q41, 26.653) (Q42, 26.241) (Q43, 25.625)};
    \addplot+[style={color=gggreen,fill=gggreen}]
        coordinates {(Q41, 13.992) (Q42, 14.011) (Q43, 13.998)};
\end{axis}
\begin{axis}[bar shift=9pt,hide axis]
    \addplot+[style={color=ppurple,fill=ppurple}]
        coordinates {(Q41, 12.393) (Q42, 9.152) (Q43, 8.339)};
    \addplot+[style={color=pppurple,fill=pppurple}]
        coordinates {(Q41, 27.915) (Q42, 29.474) (Q43, 26.604)};
\end{axis}
\end{tikzpicture}}
    \caption{Comparing the runtime performance of \frameworkname\ and the baselines for scale factor 2}
    \label{fig:laptop_sf2}
\end{figure*}

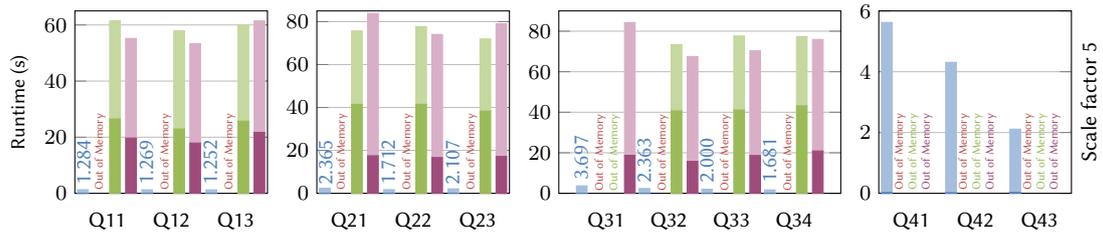
\begin{figure*}
    \subfigure{\begin{tikzpicture}[
  every axis/.style={
    ymajorgrids = true,
    major x tick style = transparent,
    xtick = data,
    enlarge x limits=0.25,
    symbolic x coords={
      Q11,
      Q12,
      Q13,
    },
    width  = 0.28*\textwidth,
    height = 4cm,
    ylabel = {Runtime (s)},
    y label style = {font=\footnotesize,at={(-0.2,0.5)}},
    ybar stacked,
    ybar=1.2pt,
    ymin=0,
    ymax=65,
    scaled y ticks = false,
    bar width=4pt,
    legend cell align=left,
    legend style={
            at={(1,1.05)},
            anchor=south east,
            column sep=1ex
    },
  },
]
\begin{axis}[bar shift=-9pt]
    \addplot[style={color=bblue,fill=bblue}]
        coordinates {(Q11, 0.015) (Q12, 0.012) (Q13, 0.012)};
    \addplot[style={color=bbblue,fill=bbblue}]
        coordinates {(Q11, 1.269) (Q12, 1.257) (Q13, 1.24)};
\end{axis}

\node[below right,rotate=90,yshift=3pt,xshift=0pt,style={bblue}] (Q11) at (0.03, 0.01) {1.284};
\node[below right,rotate=90,yshift=-20.5pt,xshift=0pt,style={bblue}] (Q12) at (0.03, 0.01) {1.269};
\node[below right,rotate=90,yshift=-44.5pt,xshift=0pt,style={bblue}] (Q13) at (0.03, 0.01) {1.252};

\node[below right,rotate=90,yshift=-4.5pt,xshift=-2pt,style={rred}] (Q21) at (0, 0.01)  {\tiny Out of Memory};
\node[below right,rotate=90,yshift=-29pt,xshift=-2pt,style={rred}] (Q22) at (0, 0.01) {\tiny Out of Memory};
\node[below right,rotate=90,yshift=-53pt,xshift=-2pt,style={rred}] (Q23) at (0, 0.01) {\tiny Out of Memory};

\begin{axis}[bar shift=3pt,hide axis]
    \addplot+[style={color=ggreen,fill=ggreen}]
        coordinates {(Q11, 26.494) (Q12, 22.913) (Q13, 25.723)};
    \addplot+[style={color=gggreen,fill=gggreen}]
        coordinates {(Q11, 34.978) (Q12, 35.017) (Q13, 34.367)};
\end{axis}
\begin{axis}[bar shift=9pt,hide axis]
    \addplot+[style={color=ppurple,fill=ppurple}]
        coordinates {(Q11, 19.712) (Q12, 17.918) (Q13, 21.709)};
    \addplot+[style={color=pppurple,fill=pppurple}]
        coordinates {(Q11, 35.471) (Q12, 35.386) (Q13, 39.74)};
\end{axis}
\end{tikzpicture}}
    \subfigure{\begin{tikzpicture}[
  every axis/.style={
    ymajorgrids = true,
    major x tick style = transparent,
    xtick = data,
    enlarge x limits=0.25,
    symbolic x coords={
      Q21,
      Q22,
      Q23,
    },
    width  = 0.28*\textwidth,
    height = 4cm,
    y label style = {font=\footnotesize,at={(-0.05,0.5)}},
    ybar stacked,
    ybar=1.2pt,
    ymin=0,
    ymax=85,
    scaled y ticks = false,
    bar width=4pt,
    legend cell align=left,
    legend style={
            at={(1,1.05)},
            anchor=south east,
            column sep=1ex
    },
  },
]
\begin{axis}[bar shift=-9pt]
    \addplot[style={color=bblue,fill=bblue}]
        coordinates {(Q21, 0.028) (Q22, 0.019) (Q23, 0.025)};
    \addplot[style={color=bbblue,fill=bbblue}]
        coordinates {(Q21, 2.337) (Q22, 1.693) (Q23, 2.082)};
\end{axis}

\node[below right,rotate=90,yshift=3pt,xshift=0pt,style={bblue}] (Q11) at (0.03, 0.01) {2.365};
\node[below right,rotate=90,yshift=-20.5pt,xshift=0pt,style={bblue}] (Q12) at (0.03, 0.01) {1.712};
\node[below right,rotate=90,yshift=-44.5pt,xshift=0pt,style={bblue}] (Q13) at (0.03, 0.01) {2.107};

\node[below right,rotate=90,yshift=-4.5pt,xshift=-2pt,style={rred}] (Q11) at (0, 0.01) {\tiny Out of Memory};
\node[below right,rotate=90,yshift=-29pt,xshift=-2pt,style={rred}] (Q12) at (0, 0.01) {\tiny Out of Memory};
\node[below right,rotate=90,yshift=-53pt,xshift=-2pt,style={rred}] (Q13) at (0, 0.01) {\tiny Out of Memory};

\begin{axis}[bar shift=3pt,hide axis]
    \addplot+[style={color=ggreen,fill=ggreen}]
        coordinates {(Q21, 41.459) (Q22, 41.476) (Q23, 38.312)};
    \addplot+[style={color=gggreen,fill=gggreen}]
        coordinates {(Q21, 34.229) (Q22, 36.236) (Q23, 33.631)};
\end{axis}
\begin{axis}[bar shift=9pt,hide axis]
    \addplot+[style={color=ppurple,fill=ppurple}]
        coordinates {(Q21, 17.501) (Q22, 16.683) (Q23, 17.164)};
    \addplot+[style={color=pppurple,fill=pppurple}]
        coordinates {(Q21, 66.327) (Q22, 57.354) (Q23, 61.863)};
\end{axis}
\end{tikzpicture}}
    \subfigure{\begin{tikzpicture}[
  every axis/.style={
    ymajorgrids = true,
    major x tick style = transparent,
    xtick = data,
    enlarge x limits=0.25,
    symbolic x coords={Q31,Q32,Q33,Q34},
    width  = 0.36*\textwidth,
    height = 4cm,
    y label style = {font=\footnotesize,at={(-0.05,0.5)}},
    ybar stacked,
    ybar=1.2pt,
    ymin=0,
    ymax=90,
    scaled y ticks = false,
    bar width=4pt,
    legend cell align=left,
    legend style={
            at={(1,1.05)},
            anchor=south east,
            column sep=1ex
    },
  },
]
\begin{axis}[bar shift=-9pt]
    \addplot[style={color=bblue,fill=bblue}]
        coordinates {(Q31, 0.031) (Q32, 0.027) (Q33, 0.032) (Q34, 0.026)};
    \addplot[style={color=bbblue,fill=bbblue}]
        coordinates {(Q31, 3.666) (Q32, 2.336) (Q33, 1.968) (Q34, 1.655)};
\end{axis}

\node[below right,rotate=90,yshift=-3pt,xshift=2pt,style={bblue}] (Q31) {3.697};
\node[below right,rotate=90,yshift=-26.5pt,xshift=1pt,style={bblue}] (Q32) {2.363};
\node[below right,rotate=90,yshift=-50pt,xshift=1pt,style={bblue}] (Q33) {2.000};
\node[below right,rotate=90,yshift=-73.5pt,xshift=1pt,style={bblue}] (Q34) {1.681};

\node[below right,rotate=90,yshift=-10.5pt,xshift=-2pt,style={rred}] (Q31a) {\tiny Out of Memory};
\node[below right,rotate=90,yshift=-34pt,xshift=-2pt,style={rred}] (Q32a) {\tiny Out of Memory};
\node[below right,rotate=90,yshift=-57.5pt,xshift=-2pt,style={rred}] (Q33a) {\tiny Out of Memory};
\node[below right,rotate=90,yshift=-81pt,xshift=-2pt,style={rred}] (Q34a) {\tiny Out of Memory};

\begin{axis}[bar shift=3pt,hide axis]
    \addplot+[style={color=ggreen,fill=ggreen}]
        coordinates {(Q31, 0) (Q32, 40.653) (Q33, 41.102) (Q34, 43.151)};
    \addplot+[style={color=gggreen,fill=gggreen}]
        coordinates {(Q31, 0) (Q32, 32.692) (Q33, 36.542) (Q34, 34.141)};
\end{axis}

\node[below right,rotate=90,yshift=-16pt,xshift=-2pt,style={ggreen}] (Q31b) {\tiny Out of Memory};

\begin{axis}[bar shift=9pt,hide axis]
    \addplot+[style={color=ppurple,fill=ppurple}]
        coordinates {(Q31, 18.75) (Q32, 15.756) (Q33, 18.656) (Q34, 20.877)};
    \addplot+[style={color=pppurple,fill=pppurple}]
        coordinates {(Q31, 65.492) (Q32, 51.741) (Q33, 51.679) (Q34, 55.039)};
\end{axis}
\end{tikzpicture}}
    \subfigure{\begin{tikzpicture}[
  every axis/.style={
    ymajorgrids = true,
    major x tick style = transparent,
    xtick = data,
    enlarge x limits=0.25,
    symbolic x coords={
      Q41,
      Q42,
      Q43,
    },
    width  = 0.28*\textwidth,
    height = 4cm,
    ylabel={Scale factor 5},
    y label style = {font=\small,at={(1.17,0.5)}},
    ybar stacked,
    ybar=1.2pt,
    ymin=0,
    ymax=6,
    scaled y ticks = false,
    bar width=4pt,
    legend cell align=left,
    legend style={
            at={(1,1.05)},
            anchor=south east,
            column sep=1ex
    },
  },
]

\begin{axis}[bar shift=-9pt]
    \addplot[style={color=bblue,fill=bblue}]
        coordinates {(Q41, 0.025) (Q42, 0.02) (Q43, 0.028)};
    \addplot[style={color=bbblue,fill=bbblue}]
        coordinates {(Q41, 5.597) (Q42, 4.288) (Q43, 2.083)};
\end{axis}

\node[below right,rotate=90,yshift=-3pt,xshift=-2pt,style={rred}] (Q11) at (0, 0.01) {\tiny Out of Memory};
\node[below right,rotate=90,yshift=-27pt,xshift=-2pt,style={rred}] (Q12) at (0, 0.01) {\tiny Out of Memory};
\node[below right,rotate=90,yshift=-51pt,xshift=-2pt,style={rred}] (Q13) at (0, 0.01) {\tiny Out of Memory};

\node[below right,rotate=90,yshift=-8pt,xshift=-2pt,style={ggreen}] (Q11a) at (0, 0.01) {\tiny Out of Memory};
\node[below right,rotate=90,yshift=-32pt,xshift=-2pt,style={ggreen}] (Q12b) at (0, 0.01) {\tiny Out of Memory};
\node[below right,rotate=90,yshift=-56pt,xshift=-2pt,style={ggreen}] (Q13c) at (0, 0.01) {\tiny Out of Memory};

\node[below right,rotate=90,yshift=-13pt,xshift=-2pt,style={ppurple}] (Q11aa) at (0, 0.01) {\tiny Out of Memory};
\node[below right,rotate=90,yshift=-37pt,xshift=-2pt,style={ppurple}] (Q12ba) at (0, 0.01) {\tiny Out of Memory};
\node[below right,rotate=90,yshift=-61pt,xshift=-2pt,style={ppurple}] (Q13ca) at (0, 0.01) {\tiny Out of Memory};

\end{tikzpicture}}
    \caption{Comparing the runtime performance of \frameworkname\ and the baselines for scale factor 5}
    \label{fig:laptop_sf5}
\end{figure*}

The runtime for the data retrieval speeds for scale factor 2 and 5 can be seen in Figures~\ref{fig:laptop_sf2} and \ref{fig:laptop_sf5}.

\subsection{Memory Usage}
\begin{figure*}
    \subfigure{\begin{tikzpicture}
    \begin{axis}[
        ymajorgrids = true,
        ytick={0,20,40,60,80,100,120},
        xtick={0,1,5,10},
        width = 0.26*\linewidth,
        height = 4cm,
        xlabel = {Scale Factor},
        ylabel = {Memory (GB)},
        ymin = 0, 
        ymax = 120,
        title = {QF 1},
        ]

        \addplot[pyCubeQ1Style]
            plot coordinates {(1, 0.088) (5, 0.088) (10, 0.088)};
            \addlegendentry{pycube\_q1}
        \addplot[pyCubeQ2Style]
            plot coordinates {(1, 0.088) (5, 0.089) (10, 0.089)};
            \addlegendentry{pycube\_q2}
        \addplot[pyCubeQ3Style]
            plot coordinates {(1, 0.089) (5, 0.089) (10, 0.088)};
            \addlegendentry{pycube\_q3}

        \addplot[JFFQ1Style]
            plot coordinates {(1, 10.335) (5, 51.429) (10, 102.769)};
            \addlegendentry{JFF\_q1}
        \addplot[JFFQ2Style]
            plot coordinates {(1, 8.694) (5, 43.226) (10, 86.367)};
            \addlegendentry{JFF\_q2}
        \addplot[JFFQ3Style]
            plot coordinates {(1, 10.335) (5, 51.429) (10, 102.769)};
            \addlegendentry{JFF\_q3}

        \addplot[JDFQ1Style]
            plot coordinates {(1, 1.961) (5, 9.377) (10, 18.661)};
            \addlegendentry{JDF\_q1}
        \addplot[JDFQ2Style]
            plot coordinates {(1, 1.961) (5, 9.377) (10, 18.661)};
            \addlegendentry{JDF\_q2}
        \addplot[JDFQ3Style]
            plot coordinates {(1, 2.189) (5, 10.436) (10, 20.716)};
            \addlegendentry{JDF\_q3}

        \addplot[SQLJQ1Style]
            plot coordinates {(1, 1.959) (5, 9.375) (10, 18.659)};
            \addlegendentry{\sql J\_q1}
        \addplot[SQLJQ2Style]
            plot coordinates {(1, 1.959) (5, 9.375) (10, 18.659)};
            \addlegendentry{\sql J\_q2}
        \addplot[SQLJQ3Style]
            plot coordinates {(1, 2.3) (5, 10.308) (10, 20.525)};
            \addlegendentry{\sql J\_q3}
        \legend{}
    \end{axis}
\end{tikzpicture}}
    \subfigure{\begin{tikzpicture}
    \begin{axis}[
        ymajorgrids = true,
        ytick={0,50,100,150,200,250},
        xtick={0,1,5,10},
        width = 0.26*\linewidth,
        height = 4cm,
        xlabel = {Scale Factor},
        ymin = 0, 
        ymax = 250,
        title = {QF 2},
        ]

        \addplot[pyCubeQ1Style]
            plot coordinates {(1, 0.09) (5, 0.09) (10, 0.09)};
            \addlegendentry{pycube\_q1}
        \addplot[pyCubeQ2Style]
            plot coordinates {(1, 0.09) (5, 0.09) (10, 0.09)};
            \addlegendentry{pycube\_q2}
        \addplot[pyCubeQ3Style]
            plot coordinates {(1, 0.09) (5, 0.09) (10, 0.09)};
            \addlegendentry{pycube\_q3}

        \addplot[JFFQ1Style]
            plot coordinates {(1, 21.389) (5, 106.654) (10, 212.959)};
            \addlegendentry{JFF\_q1}
        \addplot[JFFQ2Style]
            plot coordinates {(1, 20.545) (5, 102.435) (10, 204.758)};
            \addlegendentry{JFF\_q2}
        \addplot[JFFQ3Style]
            plot coordinates {(1, 20.545) (5, 102.435) (10, 204.758)};
            \addlegendentry{JFF\_q3}

        \addplot[JDFQ1Style]
            plot coordinates {(1, 4.099) (5, 19.71) (10, 39.151)};
            \addlegendentry{JDF\_q1}
        \addplot[JDFQ2Style]
            plot coordinates {(1, 3.818) (5, 18.248) (10, 36.283)};
            \addlegendentry{JDF\_q2}
        \addplot[JDFQ3Style]
            plot coordinates {(1, 3.817) (5, 18.248) (10, 36.283)};
            \addlegendentry{JDF\_q3}

        \addplot[SQLJQ1Style]
            plot coordinates {(1, 3.284) (5, 15.999) (10, 31.904)};
            \addlegendentry{\sql J\_q1}
        \addplot[SQLJQ2Style]
            plot coordinates {(1, 2.807) (5, 13.626) (10, 27.158)};
            \addlegendentry{\sql J\_q2}
        \addplot[SQLJQ3Style]
            plot coordinates {(1, 2.807) (5, 13.626) (10, 27.158)};
            \addlegendentry{\sql J\_q3}
        \legend{}
    \end{axis}
\end{tikzpicture}}
    \subfigure{\begin{tikzpicture}
    \begin{axis}[
        ymajorgrids = true,
        ytick={0,10,20,30,40,50,60,70,80},
        xtick={0,1,5,10},
        width = 0.25*\linewidth,
        height = 4cm,
        xlabel = {Scale Factor},
        ymin = 0, 
        ymax = 80,
        title = {QF 3},
        ]

        \addplot[pyCubeQ1Style]
            plot coordinates {(1, 0.09) (5, 0.09) (10, 0.09)};
            \addlegendentry{pycube\_q1}
        \addplot[pyCubeQ2Style]
            plot coordinates {(1, 0.09) (5, 0.09) (10, 0.09)};
            \addlegendentry{pycube\_q2}
        \addplot[pyCubeQ3Style]
            plot coordinates {(1, 0.09) (5, 0.09) (10, 0.09)};
            \addlegendentry{pycube\_q3}
        \addplot[pyCubeQ4Style]
            plot coordinates {(1, 0.091) (5, 0.09) (10, 0.09)};
            \addlegendentry{pycube\_q4}

        \addplot[JFFQ1Style]
            plot coordinates {(1, 7.709) (5, 37.863) (10, 75.557)};
            \addlegendentry{JFF\_q1}
        \addplot[JFFQ2Style]
            plot coordinates {(1, 6.208) (5, 30.363) (10, 60.561)};
            \addlegendentry{JFF\_q2}
        \addplot[JFFQ3Style]
            plot coordinates {(1, 4.661) (5, 22.629) (10, 45.095)};
            \addlegendentry{JFF\_q3}
        \addplot[JFFQ4Style]
            plot coordinates {(1, 4.943) (5, 24.035) (10, 47.907)};
            \addlegendentry{JFF\_q4}

        \addplot[JDFQ1Style]
            plot coordinates {(1, 4.427) (5, 21.481) (10, 42.816)};
            \addlegendentry{JDF\_q1}
        \addplot[JDFQ2Style]
            plot coordinates {(1, 4.192) (5, 20.307) (10, 40.451)};
            \addlegendentry{JDF\_q2}
        \addplot[JDFQ3Style]
            plot coordinates {(1, 4.192) (5, 20.307) (10, 40.451)};
            \addlegendentry{JDF\_q3}
        \addplot[JDFQ4Style]
            plot coordinates {(1, 4.192) (5, 20.307) (10, 40.451)};
            \addlegendentry{JDF\_q4}

        \addplot[SQLJQ1Style]
            plot coordinates {(1, 3.667) (5, 17.979) (10, 35.863)};
            \addlegendentry{\sql J\_q1}
        \addplot[SQLJQ2Style]
            plot coordinates {(1, 3.837) (5, 18.83) (10, 37.564)};
            \addlegendentry{\sql J\_q2}
        \addplot[SQLJQ3Style]
            plot coordinates {(1, 2.807) (5, 13.626) (10, 27.158)};
            \addlegendentry{\sql J\_q3}
        \addplot[SQLJQ4Style]
            plot coordinates {(1, 2.807) (5, 13.626) (10, 27.158)};
            \addlegendentry{\sql J\_q4}
        \legend{}
    \end{axis}
\end{tikzpicture}}
    \subfigure{\begin{tikzpicture}
    \begin{axis}[
        ymajorgrids = true,
        ytick={0,20,40,60,80,100,120,140},
        xtick={0,1,5,10},
        width = 0.24*\linewidth,
        height = 4cm,
        xlabel = {Scale Factor},
        ymin = 0, 
        ymax = 140,
        title = {QF 4},
        legend cell align = left,
        legend columns = 2,
        legend style = {
            at = {(1.05, 1)},
            anchor = north west,
            font = \tiny,
            mark options = {scale = 0.5},
            row sep = -0.5pt,
        },
        ]

        \addplot[pyCubeQ1Style]
            plot coordinates {(1, 0.09) (5, 0.09) (10, 0.09)};
            \addlegendentry{pyc\_q1}
        \addplot[pyCubeQ2Style]
            plot coordinates {(1, 0.09) (5, 0.09) (10, 0.09)};
            \addlegendentry{pyc\_q2}
        \addplot[pyCubeQ3Style]
            plot coordinates {(1, 0.091) (5, 0.091) (10, 0.091)};
            \addlegendentry{pyc\_q3}
        \addlegendimage{pyCubeQ4Style}
        \addlegendentry{pyc\_q4}

        \addplot[JFFQ1Style]
            plot coordinates {(1, 12.482) (5, 61.783) (10, 123.376)};
            \addlegendentry{JFF\_q1}
        \addplot[JFFQ2Style]
            plot coordinates {(1, 12.482) (5, 61.783) (10, 123.376)};
            \addlegendentry{JFF\_q2}
        \addplot[JFFQ3Style]
            plot coordinates {(1, 12.201) (5, 60.377) (10, 120.564)};
            \addlegendentry{JFF\_q3}
        \addlegendimage{JFFQ4Style}
        \addlegendentry{JFF\_q4}

        \addplot[JDFQ1Style]
            plot coordinates {(1, 6.946) (5, 34.284) (10, 68.269)};
            \addlegendentry{JDF\_q1}
        \addplot[JDFQ2Style]
            plot coordinates {(1, 7.136) (5, 35.231) (10, 70.145)};
            \addlegendentry{JDF\_q2}
        \addplot[JDFQ3Style]
            plot coordinates {(1, 7.178) (5, 35.423) (10, 70.583)};
            \addlegendentry{JDF\_q3}
        \addlegendimage{JDFQ4Style}
        \addlegendentry{JDF\_q4}

        \addplot[SQLJQ1Style]
            plot coordinates {(1, 4.497) (5, 22.126) (10, 44.155)};
            \addlegendentry{\sql J\_q1}
        \addplot[SQLJQ2Style]
            plot coordinates {(1, 4.599) (5, 22.639) (10, 45.18)};
            \addlegendentry{\sql J\_q2}
        \addplot[SQLJQ3Style]
            plot coordinates {(1, 4.599) (5, 22.639) (10, 45.18)};
            \addlegendentry{\sql J\_q3}
        \addlegendimage{SQLJQ4Style}
        \addlegendentry{\sql J\_q4}

    \end{axis}
\end{tikzpicture}}
    \caption{Cluster memory growth over scale factors}
    \label{fig:cluster_memory_comparison}
\end{figure*}
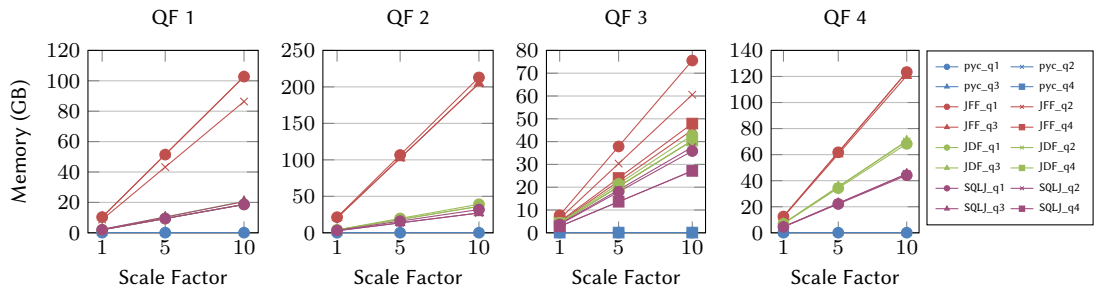

The memory usage of \frameworkname\ and the baselines on the cluster is shown in Figure~\ref{fig:cluster_memory_comparison}.
The figure shows how much memory the baselines need in order to perform the SSB queries on higher scale factors.

\subsection{Code Comparison}
SSB query 4.1 implemented in the JDF baseline is shown in Listing~\ref{lst:query41JDF}.
\begin{lstlisting}[caption={Query 4.1 of SSB using the JDF baseline},label={lst:query41JDF}]
with engine.connect() as conn:
    fact_table = pd.read_sql("lineorder", 
                             conn, 
                             columns=[
                                "lo_orderdate",
                                "lo_suppkey",
                                "lo_custkey",
                                "lo_partkey",
                                "lo_revenue",
                                "lo_supplycost"
                            ])
    date_table = pd.read_sql("date", 
                             conn, 
                             columns=[
                                 "d_datekey", 
                                 "d_monthkey"
                             ])
    month_table = pd.read_sql("month", 
                              conn, 
                              columns=[
                                "mo_monthkey", 
                                "mo_yearkey"
                              ])
    year_table = pd.read_sql("year", conn)
    part_table = pd.read_sql("part", 
                             conn, 
                             columns=[
                                "p_partkey", 
                                "p_brand1key"
                             ])
    brand_table = pd.read_sql("brand1", 
                              conn, 
                              columns=[
                                "b_brand1key", 
                                "b_categorykey"
                              ])
    category_table = pd.read_sql("category", 
                                 conn, 
                                 columns=[
                                    "ca_categorykey", 
                                    "ca_mfgrkey"
                                 ])
    mfgr_table = pd.read_sql("mfgr", conn)
    supplier_table = pd.read_sql("supplier", 
                                 conn, 
                                 columns=[
                                    "s_suppkey", 
                                    "s_citykey"
                                 ])
    customer_table = pd.read_sql("customer", 
                                 conn, 
                                 columns=[
                                    "c_custkey", 
                                    "c_citykey"
                                 ])
    city_table = pd.read_sql("city", 
                             conn, 
                             columns=[
                                "ci_citykey", 
                                "ci_nationkey"
                             ])
    nation_table = pd.read_sql("nation", conn)
    region_table = pd.read_sql("region", conn)
engine.dispose()

date1 = date_table.merge(month_table, 
                         left_on="d_monthkey", 
                         right_on="mo_monthkey")
date2 = date1.merge(year_table, 
                    left_on="mo_yearkey", 
                    right_on="y_yearkey")

part1 = part_table.merge(brand_table, 
                         left_on="p_brand1key", 
                         right_on="b_brand1key")
part2 = part1.merge(category_table, 
                    left_on="b_categorykey", 
                    right_on="ca_categorykey")
part3 = part2.merge(mfgr_table, 
                    left_on="ca_mfgrkey", 
                    right_on="m_mfgrkey")

supp_geo1 = city_table.merge(nation_table, 
                             left_on="ci_nationkey", 
                             right_on="n_nationkey")
supp_geo2 = supp_geo1.merge(region_table, 
                            left_on="n_regionkey", 
                            right_on="r_regionkey")

cust_geo1 = city_table.merge(nation_table, 
                             left_on="ci_nationkey", 
                             right_on="n_nationkey")
cust_geo2 = cust_geo1.merge(region_table, 
                            left_on="n_regionkey", 
                            right_on="r_regionkey")

supp = supplier_table.merge(supp_geo2, 
                            left_on="s_citykey", 
                            right_on="ci_citykey")

cust = customer_table.merge(cust_geo2, 
                            left_on="c_citykey", 
                            right_on="ci_citykey")

merged_table1 = fact_table.merge(date2, 
                                 left_on=
                                    "lo_orderdate", 
                                 right_on=
                                    "d_datekey")
merged_table2 = merged_table1.merge(part3, 
                                    left_on=
                                      "lo_partkey", 
                                    right_on=
                                      "p_partkey")
merged_table3 = merged_table2.merge(supp, 
                                    left_on=
                                      "lo_suppkey", 
                                    right_on=
                                      "s_suppkey")
merged_table = merged_table3.merge(cust, 
                                   left_on=
                                     "lo_custkey", 
                                   right_on=
                                     "c_custkey", 
                                   suffixes=
                                     (None, "_c"))

filtered_table = merged_table[
    (merged_table["r_region_c"] == "AMERICA")
    & (merged_table["r_region"] == "AMERICA")
    & (
            (merged_table["m_mfgr"] == "MFGR#1")
            | (merged_table["m_mfgr"] == "MFGR#2")
    )]
filtered_table["profit"] = filtered_table.apply(
                            lambda x: 
                                x.lo_revenue 
                                - x.lo_supplycost, 
                            axis=1)
return filtered_table.pivot_table(
    values="profit",
    index="n_nation_c",
    columns="y_year",
    aggfunc=np.sum
)
\end{lstlisting}

The \inlinecode{engine.dispose()} statement is included in the JDF baseline count and the \inlinecode{with} statement is counted as one statement.
The engine is created using \inlinecode{SQLAlchemy}'s \inlinecode{create_engine} method~\cite{sqlalchemy}.
The \inlinecode{with} and \inlinecode{engine.dispose()} statements manage the connection to the database.
The statements for connection management are included in the pandas count and not in the \frameworkname\ count making the comparison somewhat skewed.
However the connection management statements for \frameworkname\ only need to be done once while in pandas they need to be repeated everytime access to the database is needed.
Listing~\ref{lst:query41JDFchained} shows how the JDF baseline for query 4.1 can be shortened by chaining the \inlinecode{merge} methods together.

\begin{lstlisting}[caption={Chaining the merge methods in the JDF baseline},label={lst:query41JDFchained}]
date = date_table.merge(month_table, 
                        left_on="d_monthkey", 
                        right_on="mo_monthkey") \
                 .merge(year_table, 
                        left_on="mo_yearkey", 
                        right_on="y_yearkey")

part = part_table.merge(brand_table, 
                        left_on="p_brand1key", 
                        right_on="b_brand1key") \
                 .merge(category_table, 
                        left_on="b_categorykey", 
                        right_on="ca_categorykey") \
                 .merge(mfgr_table, 
                        left_on="ca_mfgrkey", 
                        right_on="m_mfgrkey")

supp = city_table.merge(nation_table, 
                        left_on="ci_nationkey", 
                        right_on="n_nationkey") \
                 .merge(region_table, 
                        left_on="n_regionkey", 
                        right_on="r_regionkey") \
                 .merge(supplier_table, 
                        left_on="ci_citykey", 
                        right_on="s_citykey")

cust = city_table.merge(nation_table, 
                        left_on="ci_nationkey", 
                        right_on="n_nationkey") \
                 .merge(region_table, 
                        left_on="n_regionkey", 
                        right_on="r_regionkey") \
                 .merge(customer_table, 
                        left_on="ci_citykey", 
                        right_on="c_citykey")

merged_table = fact_table.merge(date, 
                                left_on=
                                  "lo_orderdate", 
                                right_on=
                                  "d_datekey") \
                         .merge(part, 
                                left_on=
                                  "lo_partkey", 
                                right_on=
                                  "p_partkey") \
                         .merge(supp, 
                                left_on=
                                  "lo_suppkey", 
                                right_on=
                                  "s_suppkey") \
                         .merge(cust, 
                                left_on=
                                  "lo_custkey", 
                                right_on=
                                  "c_custkey", 
                                suffixes=
                                  (None, "_c"))
\end{lstlisting}

As such \frameworkname\ is much easier to use for data cube analysis than pandas.

Additionally, the JDF baseline includes a list of column names in each \inlinecode{read_sql} method.
This was done for performance so only relevant columns would be loaded into memory.
However many data scientists would usually load the entire table into memory and explore the data before filtering away unneeded columns.
Mention this in one of the previous sections
Using pandas also requires the user to know the name of all relevant keys to be used when merging dataframes.
\end{document}